\documentstyle[12pt]{article}
\topmargin - 1in
\begin{document}
\begin{flushright}
PH-205(rus) 1994

hep-th/9602003 
\end{flushright}

\begin{center}
\large{{\bf Quantum Interaction $\phi^4_4$: \\the Construction of
       Quantum Field \\defined as a Bilinear Form}
\\Edward  P. Osipov
\\            Department of Theoretical Physics
\\Institute for Mathematics
\\630090 Novosibirsk 90
\\RUSSIA}
\medskip
\\E-mail address (Relcom): osipov@math.nsk.su
      \date{}
\end{center}

\begin{abstract}

We construct the  solution
 $\phi(t,{\bf x})$ of the quantum wave equation
$\Box\phi + m^2\phi + \lambda:\!\!\phi^3\!\!:  = 0$
as a bilinear form which can be expanded over Wick polynomials
of the free $in$-field, and where
$:\!\phi^3(t,{\bf x})\!: $ is defined
as the normal ordered product with respect to
the free $in$-field.
The constructed solution is correctly
defined as a bilinear form on
 $D_{\theta}\times D_{\theta}$, where $D_{\theta}$
is a dense linear subspace in the Fock space of the free
$in$-field.
On  $D_{\theta}\times D_{\theta}$
the diagonal
Wick symbol of this bilinear form satisfies
the nonlinear classical wave equation.
\end{abstract}

\section{Introduction}

The construction of an interacting quantum field, which
satisfies a system of Wightman axioms (or a physical
and/or mathematical analogue) is a central problem in
quantum field theory.  It seems that the best starting
point would be a relativistic dynamical equation of
motion with some interpretation of nonlinear terms.  The
description of a physical vacuum as a measure on the
configuration space or on the space of trajectories is
closely connected with dynamical equations of motion and
quantum mechanics.  However here we consider a possible
description of dynamics and leave a possible description
of the vacuum for the future.

In the present paper we consider a self-interacting scalar
quantum field in four-\-di\-men\-si\-o\-n\-al
Minkowski space-time satisfying the
following relativistic wave equation
 $$ \Box\phi(t,{\bf x})
+m^2\phi(t,{\bf x})+
\lambda:\!\phi^3(t,{\bf x})\!:=0,\eqno(1.1) $$
or in the form of integral equation
 $$
\phi(t,{\bf x})=\phi_{in}(t,{\bf x})
-\lambda\int^t_{-\infty}\int
R(t-\tau,{\bf x}-{\bf y})
:\!\phi^3(\tau,{\bf y})\!:d\tau d^{3}y.
\eqno(1.2) $$

A principal barrier of this way appears as difficulties
associated with the definition of a product of fields
given at the same point.  This difficulty of the
definition of a local product is connected with a
singular dependence of the field on space-\-time
coordinates.  Usually one tries to solve this problem by
renormalization.

We consider Equations (1.1) and (1.2) and the definition
of the product of fields at a point in the following way.
First of all, we construct the solution of the quantum
Yang-Feldman equation (1.2) in the class of bilinear form
acting in the Fock Hilbert space corresponding to the
free quantum $in$-field.  In the other words, we seek the
solution of the Yang-Feldman equation (1.2), in fact, in
the form of the expansion of the solution in terms of the
creation and annihilation operators of the $in$-field.
Second, we define the product of the free $in$-field and
its normal-ordered product ( $=$ the Wick product ).

In the Fock Hilbert space any operator or any bilinear
form (belonging to a wide class of operators or bilinear
forms) can be approximated by Wick polynomials.  In the
Fock space one can define a product of bilinear forms for
bilinear forms expanded in terms of normal-ordered Wick
polynomials.  The normal-ordered functionals of the
creation and annihilation operators have a natural dense
domain of definition, $\sigma \subset {\cal H}_{in}$,
which is some analog to the Schwartz space. The are
linear continuous maps from $\sigma $ into $\sigma '$,
 where $\sigma '$ is the dual of  $\sigma$, i.e. they are
bilinear forms on $\sigma  \times \sigma,$ see
\cite{1}-\cite{3}.  In our paper we use the construction
of Wick polynomials of the free field given in
\cite{4}-\cite{11}.

We solve the Yang-Feldman equation (1.2) by constructing
the expansion of the quantum field in terms of Wick
polynomials of the free $in$-field.

Since this expansion converges not on all vectors, so to
construct the quantum field as a bilinear form we choose
the special subspace $D_{\theta}$ of vectors in the Fock
Hilbert space. Namely, we take coherent vectors near to
the vacuum of the $in$-field and their finite linear
combinations.  Here a coherent vectors near to the vacuum
means the vector of the form $|z\rangle=
 \exp(za^+_{in})\Omega$ with small complex-valued test
function $z$ ( $\Omega$ is the vacuum vector, $z\in {\cal
S}$ and has a small $F$-norm, the definition of the
 $F$-norm see in Sect. 2 ).  We note that  $D_{\theta}$
is dense in the Fock space.

The considered expansion of the quantum field in terms of
the Wick polynomials of the free $in$-field converges on
the coherent vectors near to the vacuum and defines the
solution of the quantum wave equation (1.1), (1.2) as a
 bilinear form on $D_{\theta}\times D_{\theta}.$

The considered construction uses, in fact, the idea that
the creation operator is conjugate to the annihilation
one, moreover every creation vector is an eigenvector of
all annihilation operators. In other words, matrix
 elements of quantum field on coherent vectors, i.e. Wick
symbols, are solutions of classical Yang-Feldman wave
equation with complex $in$-data.

However, a complication arises here. This complication is
 connected with the existence and the construction of
complex solutions of classical (real) wave equation. We
overcome this complication by using coherent vectors near
to the vacuum. To coherent vectors near to the vacuum
 correspond small matrix elements of the quantum field
and small complex $in$-data of classical wave equation.
 The convergence of solutions of the classical wave
equation for small complex initial $in$-data gives us the
convergence of the considered expansion and allows us to
construct the quantum field as a bilinear form defined on
the subspace, generated by linear combinations of
coherent vectors corresponding to small complex
$in$-data. Therefore, we construct the bilinear form by
using its Wick symbols for coherent vectors near to the
vacuum only.

The same consideration
allows to construct the bilinear form
 $\phi_{out}$  corresponding to the quantum
 $out$-field.

The constructed quantum field is a scalar with respect to
the Poincar\'e transformation
$$
U(a,\Lambda)\phi(t,{\bf x}) U(a,\Lambda)^{-1}
=\phi((a,\Lambda)(t,{\bf x})),\quad\mbox{ where }
U(a,\Lambda)=U_{in}(a,\Lambda).
$$
 The generator of the translation subgroup
(that is, the Hamiltonian
and the momentum operator) satisfy the spectrum
condition.
The  constructed  field is non-local with respect
to the free
 $\phi_{in}(t,{\bf x})$ field.
The question about locality of the
  constructed  field is open.
This question is closely connected with
 the question about
a  structure of
the  bilinear form and
with the question about the existence of a measure
corresponding to the vacuum and about its support.  It
would be interesting to represent the constructed
bilinear form $\phi(t,x)$ as an  operator-valued
generalized function or as an operator-valued
hyperfunction.

In conclusion, we remark that the considered construction
has been suggested by Heifets  \cite{12}. He also
constructed the quantum field with the help of small
complex initial data, however Heifets \cite{12} used
 instead of  $F$-norm more complicated variant of
 $R$-norm.

R\c{a}czka \cite{13} also tried to construct the quantum
field as a bilinear form. He used for the construction a
unproved assumption about the Wick symbols of
approximations and their convergence to the solution of
(real) wave equation for any (not necessary small)
complex $in$-data. This assumption is incorrect in
general.  We go around this difficulty considering small
complex initial $in$-data and extending the results of
 Morawetz and Strauss \cite{14,15}, for this case, see
 also \cite{16}-\cite{18}.

Our consideration is the following. In Sect. 2 we
formulate and prove the assertions (Theorems 2.3 and 2.4)
that we need for solutions of non-linear (real) wave
equation with small complex $in$-data. In Sect. 3 we
prove some estimates for the non-linear part of the
classical wave equation (Lemmas 2.1 and 2.2). In Sect. 4
we describe field as a bilinear form (Theorems 4.1-4.3)
and in Sect. 5 we discuss the obtained results and its
connection with other approaches.

\section{Solution of the wave equation for
\newline small complex initial $in$-data}
In this section we consider global complex solutions of
classical (real) nonlinear wave equation
$$
u_{tt}-\Delta u+m^2 u+\lambda u^3=0,\;\;\;\; m>0,\;
\lambda>0.\eqno(2.1) $$
To construct the hermitian (scalar) quantum field we need
the solutions of (2.1) for small complex initial
 $in$-data.
First we rewrite the equation (2.1) in the integral form
$$
u(t,{\bf x})=u_T(t,{\bf x})-\lambda\int^t_T\int
R(t-\tau,{\bf x}-{\bf y})u^3(\tau,{\bf y})d\tau
d^3y . \eqno(2.2) $$
Here $u_T(t,{\bf x})$
is the complex solution of the free equation
that corresponds to the complex Cauchy data at the time
 $T$, $R(t,{\bf x})$  is the Riemann function
of the linear equation, i.e. the free solution with
Cauchy data
$R(0,{\bf x})=0,\;\; R_t(0,{\bf x})= \delta({\bf x}).$
We remark that
$$R(t,{\bf x})=-R(-t,{\bf x}), \quad R(t,{\bf x})=
{\sin(t(-\Delta+m^2)^{1\over2})
\over(-\Delta+m^2)^{1\over2}}({\bf x})$$
and for $t>0$
$$ R(t,{\bf x})
={\delta(t-|{\bf x}|)\over 4\pi \, t}+{1\over
4\pi}\theta(t-|{\bf x}|)\, m
J_1(m(t^2-|{\bf x}|^2)^{1/2})
\,(t^2-|{\bf x}|^2)^{-1/2}. $$
To construct the quantum field
we need the solutions of the
equation $$ u(t,{\bf x})=u_{
in}(t,{\bf x})-\lambda\int^t_{-\infty}
\int R(t-\tau,{\bf x}-{\bf y})
u^3(\tau,{\bf y})
d\tau d^3y\eqno(2.3) $$
for small complex $in$-data.
Here  $u_{in}$ is a solution
of the free equation, corresponding to
complex $in$-data.

To construct the solutions of the Yang-Feldman
equation (2.3) we extend Morawetz's and  Strauss'
results \cite{14}
on the case of small complex-valued $in$-data.

In order to describe our results we define the
energy norm
$$
\|u(t)\|^2_e=\int(|u_t(t,{\bf x})|^2+|\nabla
u(t,{\bf x})|^2+m^2|u(t,{\bf x})|^2)d^3x $$
and the $F$-norm \cite{14} $$
\|u\|^2_F=\sup_{t}\|u(t)\|^2_e+\sup_{t}\sup_{{\bf x}}
|u(t,{\bf x})|^2
+\int^\infty_{-\infty}\sup_{{\bf x}}
|u(t,{\bf x})|^2d t \eqno(2.4) $$
for the Banach space $F^C$ of continuous complex-valued
functions with finite $F$-norm.

Define the Banach space
${\cal F}^C$
of complex-valued solutions of the free equation
 $$ u_{tt}-\Delta u+m^2 u=0, $$
which is a subspace of the Banach space
$F^C.$
For this purpose we define the subspace
 ${\cal F}^C_1$
as the space of the free solutions whose Cauchy data
$(\varphi({\bf
x}),\pi({\bf x}))$
at the zero time
are such that  $\varphi({\bf
x})$ with its first and second derivatives
belong  to
$L^1\cap L^2$, and the third derivatives
are in
$L^1$, $\pi({\bf
x})$ with its first derivatives are in
 $L^1\cap L^2$ and second derivatives are in  $L^1.$
A solution of the free equation with initial data
from ${\cal F}^C_1$
has the uniform decay like
 $(1+|t|)^{-3/2}$ for $t\to\infty$ and has a finite
$F$-norm (a complex free solution can be considered
similar to the real one, see Appendix B \cite{14}).
Define  ${\cal F}^C$
as the completion of ${\cal F}^C_1$ in
the  $F$-norm. It is clear that ${\cal F}^C$
is a closed subspace of $F^C.$

Let  ${\cal F}$ and  $F$ denote the corresponding
real subspaces of the spaces ${\cal F}^C$ and $F^C.$

To construct a solution for small complex $in$-data we
formulate two lemmas, which we prove in Sect. 3.

To formulate these lemmas we introduce the notations
that we need.

Let $M=[a,b], \; -\infty\leq a<b\leq +\infty$, and in
addition
$M=M_1\cup M_2$, where $M_1 = [a,b]  \cap \\ \{\tau\mid
|t-\tau|\geq\delta\},$ $M_2 =[a,b]\cap\{\tau\mid
|t-\tau|<\delta\},$ $\delta>0.$

 We define
$$[[u]]_M=\sup_{t\in M}\sup_{{\bf x}}
|u(t,{\bf x})|+\{\int_M\sup_{{\bf x}}|u(t,{\bf x})|^2
dt\}^{1/2}$$
and
$$[[u]]=[[u]]_{(-\infty,\infty)}.$$

   We introduce the norm
$$\langle u\rangle_M
=\biggl\{\sup_K\int_{K\cap M}|u|^2d{\tilde
S}\biggr\}^{1/2},\;\;\; \quad
\quad \langle u\rangle\equiv \langle
u\rangle_{(-\infty,\infty)},$$ where
$d{\tilde S}$ denotes the measure
on the surface of the cone $K$ and
 $K$ runs over all forward and backward light cones
in space-time.
On the surface of the forward or
backward light cone $K=\{(t,{\bf x})\in K \;\vert
\; t^2-|{\bf x}|^2\geq 0, \,\pm t\geq 0\}$  we
use the measure
$d\tilde S$, where  $d\tilde S=
\theta(\pm t)\delta(|t|-|{\bf x}|)dt d^3x
=\theta (\pm t)dtd S$ and  $d S=t^2
d\omega, |\omega|=1$ is the sphere measure on the
sphere of radius
$t$
in ${I\!\!R}^3$.

\medskip
Lemma 2.1.

Let
$$
I(t,{\bf x})=\int_{M_1}\int R(t-\tau,{\bf x}-{\bf y})
u(\tau,{\bf y}) v(\tau,{\bf y}) w(\tau,{\bf y})
d\tau d^3 y, $$
where $u$, $v$, and $w$
are arbitrary smooth complex-valued functions.

Then
$$
|I(t,{\bf x})|^2\leq
c(\delta,\alpha)\langle
w\rangle_M\int_{M_1}\bigl[\|u(\tau)\|^2_\infty
\|v(\tau)\|_e+\|u(\tau)\|_\infty\|v(\tau)\|_\infty
\|u(\tau)\|_e\bigr]$$
$$\|v(\tau)\|^{1-2\alpha}_2\,\|v(\tau)\|^{2\alpha}_\infty
\,\,|t-\tau|^{-3/2+3\alpha} d\tau.
$$
Here  $0\leq\alpha\leq 1/2$, $c(\delta,\alpha)$
is a constant,
depending on   $\delta$ and $\alpha$ only
(and, maybe, on the mass
which enters into the Riemann function).

\medskip
Lemma 2.2.

Let $u(t,{\bf x})$ and $v(t,{\bf x})$
be a pair of  arbitrary smooth
complex-valued functions.
Let
$${\cal R}u(t,{\bf x})=
\int _M\int R(t-\tau,{\bf x}-{\bf y})
u^3(\tau,{\bf y})d\tau d^3y. $$

(a) For any $0\leq\alpha<1/6$, we have
$$ \|{\cal
R}u\|_F \leq
c[[u]]^{1+\alpha}_M \,\|u\|^{1-\alpha}_{F,M}\,
\biggl [\|u\|_{F,M} +\langle u\rangle_M \biggr] .$$

\smallskip
(b) $$  \|{\cal R}u-{\cal
R}v\|_F
 \leq  c\biggl([[u]]_M+[[v]]_M\biggr)^{1/2}
\biggl(\|u\|_{F,M}+\langle u\rangle_M+\|v\|_{F,M}+
\langle v\rangle_M\biggl)^{3/2} $$
  $$\|u-v\|^{1/2}_{F,M}
\,\sup_{t\in M}\|u(t)-v(t)\|^{1/2}_e .$$

\medskip
Remark. The constants entering into Lemma 2.2
do not depend of $M.$

\medskip

For a solution
$u=u(t,{\bf x})$ of Equation (2.1) we denote
by $u_T=u_T(t,{\bf x})$ the free solution whose Cauchy data
at  $t=T$ agree with that of  $u$,
  $$u_T(T,{\bf x}) = u(T,{\bf x}),
\quad\quad\quad{\partial {u_T}\over
{\partial t}}(T,{\bf x})=
{\partial {u}\over{\partial t}}(T,{\bf x}),$$
and formulate now the theorems
about solutions of Equations (2.2)-(2.3)
for small complex initial data and $in$-data.

\medskip
Theorem 2.3 (Cauchy problem).

There exists a strictly positive $\theta$ such that for any
$S$, $-\infty<S<\infty,$ and $u_S\in {\cal F}^C,
\|u_S\|_F<\theta$, there exists a unique global solution
 $u$ of Equation (2.2) with finite $F$-norm and whose
Cauchy data
 at time $S$ equal that of  $u_S.$  In addition
$\|u\|_F< 2\theta$.
The free solution  $u_T$, whose Cauchy data at time
 $T$ equal that of $u,$ also belongs to ${\cal F}^C$.
Furthermore, $u_T$ depends continuously on
 $u_S$ in  ${\cal F}^C$ and  $\|u_T\|_F< 2\theta$.

There exists  a unique free solution  $u_{in}$ and
 a unique free solution  $u_{out}$  such that
  $$\|u_{in}(t)-u(t)\|_e \to 0 \; \mbox{ for }\;
t\to -\infty \; \mbox{ and }\; \|u_{out}(t)-u(t)\|_e \to 0
\; \mbox{ for }\; t\to +\infty,$$  in this case $u_{in},
u_{out}\in{\cal F}^C$ and the  $F$-norm of $u_{in}$  and
of
$u_{out}$ is less than  $2\theta$.

\medskip
Theorem 2.4 (Cauchy problem at $t=-\infty$).

There exists a strictly positive $\theta$ such that for
$u_{in}\in{\cal F}^C$ and  $\|u_{in}\|_F<\theta$ there
exists
 a unique global solution
 $u$ of Equation (2.3) with finite $F$-norm and which
converges
in the energy norm to  $u_{in}$ for  $t\to -\infty$.
In addition  $\|u\|_F< 2\theta$.
For any
$T$, $-\infty <T<+\infty$, the free solution $u_T,$
whose Cauchy data
at time  $t=T$ equal that of $u,$
also belongs to ${\cal F}^C$
and in this case
 $\|u_T\|_F< 2\theta$.
There exists
 a unique free solution
 $u_{out}$ such that  $\|u(t)-u_{out}(t)\|_e\to
0$ for $t\to +\infty$,
in addition $\|u_{out}\|_F< 2\theta$.
 Furthermore,
 $u_T, u_{out}$
 depend continuously on
 $u_{in}$ in ${\cal F}^C$.
 $u(t,{\bf x})$ also depends
continuously on $u_{in}$ in ${\cal F}^C$.

\medskip
Remarks.

1) $\theta$  depends on the mass and
the coupling constant
in the non-linear equation (2.1) and its
``smallness'' depends only on
the value of constants in the bounds of
Lemmas 2.1 and  2.2.

2) If  $u^3$ is replaced by  $F(u)$ in
Equation  (2.3), this theorem
depends only on the property
$|F''(u)|={\it O}(|u|)$ as $u\to 0.$

\medskip
Proof of Theorems 2.3 and 2.4.

The proof of Theorem 2.3 is completely
the same as the proof of Theorem
2.4. Therefore, we restrict ourselves
to the proof of Theorem 2.4.

 First consider the uniqueness of a solution of the
Cauchy problem at
$t=-\infty$.  For the  $u$  with the initial data
$u_{in}$
from ${\cal F}^C_\theta\equiv
{\cal F}^C\cap\{u\in{\cal
F}^C \vert \;\|u\|_F\leq\theta\}$
and with finite  $F$-norm,
$\|u\|_F\leq\theta$, there is the representation
$$
u_{in}(t)=u(t)+\lambda\int ^t_{-\infty} R(t-\tau)\ast
u^3(\tau)d\tau.
$$
If $u$ and  $v$ are two solutions as in the statement
of Theorem 2.4, then
 $$ u(t)-v(t)=-\lambda\int ^t_{-\infty}
R(t-\tau)\ast [u^3(\tau)-v^3(\tau)]d\tau.
$$
Taking the energy norm, we get
$$
\sup_{t\leq S}\|u(t)-v(t)\|_e\leq{\lambda\over
m}\,\sup_{\tau\leq S}\|u(\tau)-v(\tau)\|_e \int
^S_{-\infty}(\|u(\tau)\|_\infty+\|v(\tau)\|_\infty)^2d\tau,
$$
or
$$
\sup_t\|u(t)-v(t)\|_e\leq{\lambda\over
m}\,\sup_t\|u(t)-v(t)\|_e \int
^\infty _{-\infty}(\|u(\tau)\|_\infty
+\|v(\tau)\|_\infty)^2d\tau .
$$
Since
$${\lambda\over m}\int ^\infty
 _{-\infty}(\|u(\tau)\|_\infty
+\|v(\tau)\|_\infty)^2 d\tau
\leq {\lambda\over m}
\,(\|u\|_F+\|v\|_F)^2 \leq {4\lambda\theta
^2\over m},$$
choosing
 $\;\; 4\lambda\theta ^2/ m^2 < 1$
we get  $$\sup_t\|u(t)-v(t)\|_e=0 $$
and, since  $u(t)$ and  $v(t)$  belong to
$F^C$ and, thus, are continuous,
 $u(t)=v(t).$ This proves the uniqueness.

To construct the solution  $u,$ we shall
solve the integral equation.
Since we shall require  $u_T\in{\cal F}^C$, the
construction must exhibit $u$ as the limit of smooth
  functions. For this reason, we first solve the ordinary
Cauchy problem with initial data at a time $S.$ Let
 $u_S\in {\cal F}^C$ be a free solution with complex
$C^2$ data of compact support.

Define
$$
{\cal R}u(t,{\bf x})=-\int ^t_S R(t-\tau)\ast
u^3(\tau)d\tau.
$$
We solve the equation
$$
u=u_T+\lambda{\cal R}u
$$
by successive approximations:
$$
u^{(0)}=u_S,\;
u^{(n)}=u_S+\lambda{\cal R}u^{(n-1)},\;\;
n=1,2,...\;  .  $$
Each $u^{(n)}$ is of class  $C^2$ because
$u_S$ is.

We claim, that
 $\theta$ so small that if
$\|u_S\|_F<\theta$, then
\begin{tabbing}
\hspace{1in}\=\hspace{1in} \=\quad \kill
\>$(i)$ \=\quad $\|u^{(n)}\|_F\leq\theta,$\\
\>$(ii)$ \=\quad $\langle u^{(n)}
\rangle\leq{{2\theta}/m}$
\end{tabbing}
for all $n=0,1,2,...\; .$

We prove the claim by induction
on $n$. If $n=0$,  $(i)$
is true by definition. The estimate $(ii)$ follows
from the inequality
$(i)$ and the simple energy inequality
 $\langle
u_S\rangle\leq{2\over m}\|u_S\|_e.$
This energy inequality can be proved
with the help of relation
(2.7) for the energy-momentum density
(2.8)-(2.9) in the same way
as the energy inequality (2.10). We note that
$u^{(0)}=u_S$ is the solution of the
free equation and for
$u_S$ the right side of (2.10) is equal to zero.

Next, we have
$$\|u^{(n)}-u_S\|_F
\leq \mbox{ (by Lemma 2.2(a)) }
 c \|u^{(n-1)}\|^2_F \,\Bigl(\|u^{(n-1)}\|_F+\langle
u^{(n-1)}\rangle\Bigr)$$
 $$\leq \mbox{ (by the induction assumption) }
 \lambda c (1+{2\over m})\theta ^3.\eqno(2.5)
$$
Choose $\theta$ so small that $$\lambda c
\,(1+{2\over m})\,\theta
^3\leq{1\over 4}\theta.$$
With this choice of  $\theta$ we then have
$(i).$

To prove  $(ii)$, note that  $u^{(n)}$
is a solution of
  $$ u^{(n)}_{tt}-\Delta u^{(n)}+m^2
u^{(n)} = -\lambda(u^{(n-1)})^3\eqno(2.6)$$
and therefore enjoys the
 energy inequality.
To prove  the energy inequality we use the following
identity
$$
(u_{tt}-\Delta u+m^2 u)u_{t}^\ast
+(u_{tt}^\ast-\Delta u^\ast+m^2
u^\ast)u_{t} = {\partial E\over\partial t}+\mbox{ div }
{\bf P},\eqno(2.7)
$$
where
$$
E(t,{\bf x})=|u_{t}(t,{\bf x})|^2+|\nabla u(t,{\bf
x})|^2+m^2|u(t,{\bf x})|^2,\eqno(2.8)
$$
$$
{\bf P}(t,{\bf x})=-\nabla u(t,{\bf x}) u_{t}^\ast(t,
{\bf x})-\nabla
u^\ast(t,{\bf x})u_{t}(t,{\bf x}).\eqno(2.9)
$$
This identity is fulfilled for any (smooth) function
 $u$.
Note that
$(E(t,{\bf x}), {\bf P}(t,{\bf
x}))$ is given by the components $(T_{00}(t,{\bf x}),
T_{0i}(t,{\bf x}))$
of the energy-\-momentum tensor, see
\cite[\S{XI.14}, Addition]{19},
\cite[ch.1, \S{2.2}, p.23]{20}, see also
\cite[Theorem 2.1]{21}, and is the
 energy-momentum density of the
 complex field and not of a real one)
To obtain the energy inequality
we multiply the equation (2.6) by $u^{(n)\ast}_t$, add
the conjugate term and integrate over the part
 $K_1$ of the forward or backward light cone.
The equality  (2.6) implies that the integral over the
4-dimensional
divergence (the right side of (2.7) )
is not greater than the right side of (2.10).
On the other hand, by the Gauss theorem the integral over the
4-dimensional divergence is equal to the energy-momentum flow
(2.8)-(2.9) through the chosen part of the forward or backward
light cone and is estimated from below by the left side
of (2.10).
This gives   $$ 2^{-1/2} m^2\int_{K_1}|u^{(n)}|^2d{\tilde S}
\,-\, \sup_t\|u^{(n)}(t)\|^2_e
\leq2\lambda\int\int|(u^{(n-1)})^3u^{(n)\ast}_t|
dt d^3x, \eqno(2.10)$$
and finally we receive
$$ 2^{-1/2} m^2\langle
u^{(n)}\rangle^2 \,- \,\sup_t\|u^{(n)}(t)\|^2_e
\leq2\lambda\int\int|(u^{(n-1)})^3u^{(n)\ast}_t|
dt d^3x.\eqno(2.11)$$
Next, the right side of  (2.11) is less than
 $$ 2\lambda\sup\|u^{(n)}_t\|_2 \,\|u^{(n-1)}\|_2 \int ^\infty
_{-\infty}\|u^{(n-1)}(t)\|^2_\infty d t.\eqno(2.12) $$
 (2.5), the choice of $\theta$, and the induction
assumption imply that
 (2.12) is not greater than
 $2\lambda (\|u_S\|_F+{1\over 4}\theta) \theta
^3\leq {5\over 2}\lambda\theta ^4,$   that is
$$
\langle u^{(n)}\rangle^2\leq 2^{1/2} m^{-2} (\theta ^2+{5\over
2}\lambda\theta ^4) \leq {{4\theta^2}/{ m^2}}
  $$ for $2^{1/2}( 1 + {5\over 2}\lambda\theta ^2)\leq 4.$
 This proves  $(ii).$

\medskip
Next, we apply Lemma 2.2(b) to the difference
$$
u^{(n)}-u^{(n-1)}=\lambda{\cal R}u^{(n-1)}-\lambda{\cal
R}u^{(n-2)}.
$$
Using  $(i)-(ii),$ we obtain
$$
\|u^{(n)}-u^{(n-1)}\|_F\leq c\lambda\theta ^2(2+{4\over
m})^2\,\|u^{(n-1)}-u^{(n-2)}\|_F.
$$
Choosing the coefficient on the
right to be less than $1/2,$
through choice of $\theta,$   $\{u^{(n)}\}$
becomes a Cauchy sequence in the  $F$-norm.
Its limit is the solution  $(I-{\cal R})^{-1}u_T.$
Furthermore, this solution is a  $\;C^2$ function
 (if $u_S$
is ) as a consequence of the estimate
$$ \|Du^{(n)}(t)\|_\infty\leq\|D
u_S(t)\|_\infty+c\int ^t_S\|D
u^{(n-1)}(\tau)\|_\infty d\tau,
$$
where $D$ is a first or second derivative.

If  $u_S$ has compact support in space,
so does the solution.
 This follows from the explicit form of
the approximation $u^{(n)},$ from the fact, that
$$
\mbox{ supp } u^{(n)}(t,\cdot)\subset\{{\bf
x}\in {I\!\!R}^3 | \mbox{ dist}\Bigl({\bf x},
\mbox{supp}\bigl(u_S(S,\cdot),
{\partial {u_S}\over{\partial t}}(S,\cdot)\bigr)\Bigr)
\leq|t-S|\}
$$
and from the convergence of the series
$u=\sum_n(u^{(n+1)}-u^{(n)}).$
Of course, these statements are valid  for every $t,$
$\;-\infty<t<\infty.$

The convergence of the series
$u=\sum _n(u^{(n+1)}-u^{(n)})$ and the restrictions on
$\theta$ imply that $\|u\|_F<2\theta.$
Moreover, the convergence
of $u^{(n)}$ to $u$ in the  $F$-norm implies that
$\langle \cdot\rangle$-norm of  $u$ is
also restricted by
 ${2\theta}/ m.$  This follows from the continuity of
 $u^{(n)}$ and  $u$ and the convergence and the uniform
boundedness of the integral of
  $u^{(n)2}$ taken over the bounded part of the cone.

Now let $u_{in,k}$ be a sequence of  $\;C^2$ smooth
free solutions
of compact support which tends to
$u_{in}$ in ${\cal F}^C,$  $\|u_{in}\|_F<\theta.$
It is clear that there exists a sequence
of such $u_{in,k}.$ Let
$u_k$ be the constructed solution of
 (2.2) whose Cauchy data at time  $t=-k$
equals that of $u_{in,k}$  at time  $t=-k.$ The limit of
$u_k$ as $k\to\infty$ will be the required solution.
To prove the convergence of  $u_k$ we consider
the difference
$u_k(t)-u_l(t).$  For $k>l$
$$
u_k(t)-u_l(t)=(u_{in,k}(t)-u_{in,l}(t))-\lambda\int ^t_{-l}
R(t-\tau)\ast(u^3_k(\tau)-u^3_l(\tau)) d\tau$$
     $$-\lambda\int ^{-l}_{-k}
R(t-\tau)\ast u^3_k(\tau)d \tau.\eqno(2.13)
$$
Consider the $F$-norm. The $F$-norm of the first term goes
to zero. Lemma 2.2(b) and the uniform estimate
 $\langle u_k\rangle\leq{{2\theta}/ m}$  imply that
the $F$-norm of the second term is less than
$c\lambda\theta ^2(2+{4\over
m})^2\|u_k-u_l\|_F<{1\over 2}\|u_k-u_l\|_F.$
The estimate of this term we carry over on the left side.

The final term on the right is estimated as follows.
Let $\varepsilon>0.$ Since
 $u_{in,k}$ converges in ${\cal F}^C$, there exists an
 $L=L(\varepsilon)$ such that,
$$
[[u_{in,k}]]_{(-\infty,-L]}
=\sup _{t\leq-L}\|u_{in,k}(t)\|_\infty+
\biggl\{\int
^{-L}_{-\infty}\|u_{in,k}(t)\|^2_\infty d t\biggr\}^{1/2}
$$
$$
\leq\sup_{t\leq-L}\|u_{in}(t)\|_\infty+
\biggl\{\int
^{-L}_{-\infty}\|u_{in}(t)\|^2_\infty
dt\biggr\}^{1/2}+2\|u_{in}-u_{in,k}\|_F<\varepsilon
$$
for all $k\geq L.$

Lemma 2.2(a) and the equality
$u_k=u_{in,k}+\lambda{\cal R}u_k$ imply that
$$
[[u_k]]_{(-\infty,-L]}
\leq[[u_{in,k}]]_{(-\infty,-L]}+c\lambda\theta^2
(1+{2\over m})\;[[u_k]]_{(-\infty,-L]} \,.
$$
For sufficiently small
$\;\theta\;$ $\;c\lambda\theta^2(1+{2\over m})<{1\over
2}.$ So $[[u_k]]_{(-\infty,-L]}<2\varepsilon.$
Therefore, these arguments, the uniform estimate
$\langle u_k\rangle\leq 2\theta/m$  and
Lemma 2.2(a) imply that
the $F$-norms of the last term of
 (2.13) are not greater than
 $$c\lambda\theta^2 (1+{2\over m})
\;[[u_k]]_{(-\infty,-L]}\leq\varepsilon .$$
It follows from these estimate that
$\{u_k\}$ is a Cauchy sequence in the  $F$-norm.

Call the limit  $u.$ By passage to the limit we obtain
 $$u(t) = u_{in}(t)-\lambda\int ^t_{-\infty}R(t-\tau)\ast
u^3(\tau)d\tau, $$     whence   $$
\|u(t)-u_{in}(t)\|_e
\to 0 \quad \mbox{ as}\quad t\to
-\infty,$$  This means that we have constructed
the solution $u(t)$ for the initial $in$-data
$u_{in}(t).$

\medskip
Now let  $u_T$ be the free solution
with the Cauchy data at $t=T$
equal to the Cauchy data of $u.$ Let
 $u_k$ be defined as stated above, that is,
it is a solution of (2.2), with Cauchy data at time
 $t=-k$
equal to the Cauchy data of $u_{in,k}$ at time
$t=-k.$ Let $u_{k,T}$
be the free solution
with the Cauchy data at  $t=T$ equal to
the Cauchy data of $u_k.$ Then  $u_{k,T}$ is a
smooth free solution
given by $$ u_{k,T}=u_k(t)-\lambda\int^T_{-m}
R(t-\tau)\ast u^3_k(\tau)d\tau.  $$

Just as in the above argument, the right side
converges in ${\cal F}^C.$
If the limit of $u_{k,T}$ in ${\cal F}^C$ is
called $v,$ then
$v\in{\cal F}^C$ and $$
v(t)=u_{in}(t)-\lambda\int^T_{-\infty}(R(t-\tau)\ast
u^3(\tau)d\tau.  $$ Since the Cauchy data of
the free solution
 $v$
at $t=T$ agree with those of $u,$  $v=u_T,$ that is,  $$
u_T(t)=u_{in}(t)-\lambda\int^T_{-\infty} R(t-\tau)
\ast u^3(\tau)d\tau.
$$
The continuous dependence of $u_T$ in the $F$-norm is a
consequence of
the construction: $u$ depends continuously on  $u_{in},$
and $u_T$
on $u_{in}.$

Now we construct the solution $u_{out}$.  $u_{out}$ has
been defined
as a unique free solution such that
$\|u_{out}(t)-u(t)\|_e\to 0$ for $t\to+\infty.$
We claim that  $u_{out}$
is given by the following formula
 $$
u_{out}(t)
= u_{in}(t)-\lambda\int^{+\infty}_{-\infty}R(t-\tau)\ast
u^3(\tau)d\tau, $$
that means that the formula
 $$
u_{out}(t)=u(t)-\lambda\int^{+\infty}_t R(t-\tau)\ast
u^3(\tau)d\tau.
$$
is valid also. Indeed, the right sides are defined correctly,
have finite energy and
$u(t)$ converges to $u_{out}$
in the energy norm as $t\to +\infty.$
Direct differentiation, in the weak sense, shows that
$u_{out}$ is a free solution, so that it must coincide with
$u_{out}$. We need to show
that not only
the $F$-norm of $u_{out}$ is finite, but that
 $u_{out}\in{\cal F}^C.$

To prove the statement that $u_{out}\in{\cal F}^C$
we approximate
 $u_{in}$ by smooth solutions $u_{in,k}$ with compact support
and as  $u_k$  we take the solution of (2.2),
whose Cauchy data at time $t=-k$ agree with that of
$u_{in,k}$  at time  $t=-k.$ Then let
$u_{out,k}$ be the free solution whose Cauchy data at time
 $t=k$ agree with the Cauchy data of $u_k$ at time $t=k.$

Of course the Cauchy data of $u_k$ at any time are smooth
and of compact support.

We have the integral representation
$$
u_{out,k}(t)=u_{in,k}(t)-\lambda\int ^k_{-k} R(t-\tau)\ast
u^3_k(\tau)d\tau,
$$
whence, for  $k>l$ we have
$$
u_{out,k}(t)-u_{out,l}(t)
=(u_{in,k}(t)-u_{in,l}(t))-\lambda\int^l_{-l}
R(t-\tau)\ast\bigl(u^3_k(\tau)-u^3_l(\tau)\bigr)d\tau
$$
$$
-\lambda\int^{-l}_{-k}R(t-\tau)\ast
u^3_k(\tau)d\tau-\lambda\int^k_l
R(t-\tau)\ast u^3_k(\tau)d\tau. \eqno(2.14)
$$

Consider  the $F$-norm  of the four terms on the right as
$k, l\to\infty.$ The $F$-norm of the first term goes to
zero by assumption, the second term is less than
 $$c\lambda\theta^2(2+{4\over m})^2\|u_k-u_l\|_F<{1\over
2}\|u_k-u_l\|_F.$$ The latter two terms can be estimated
analogously to the similar term of (2.13). Let
 $\varepsilon>0.$ There exists $L=L(\varepsilon)$ such
that $$ {[[u_{in,k}]]_{(-\infty,-L]\cup[L,\infty)}
\leq\sup_{|t|\geq L}\|u_{in}(t)\|_\infty + \biggl\{
\int_{|t|\geq L}\|u_{in}(t)\|^2_\infty dt\biggr\}^{1/2} +
2\|u_{in}-u_{in,k}\|_F<\varepsilon} $$ for all $k\geq L.$

By Lemma 2.2(a) $$
[[u_k]]_{(-\infty,-L]}+[[u_k]]_{[L,\infty)} \;\leq\;
[[u_{in,k}]]_ {(-\infty,-L]}+[[u_{in,k}]]_{[L,\infty)} $$
$$ +c\lambda\theta^2(1+{2\over m})
\,[[u_k]]_{(-\infty,-L)}+c\lambda\theta^2(1+{2\over m})
\,[[u_k]]_{[L,\infty)}.  $$ For sufficiently small
$\theta\;\;\; c\lambda\theta^2(1+{2\over m})<1/2.$ So
$$[[u_k]]_{(-\infty,-L]}+[[u_k]]_{[L,\infty)}<2\varepsilon.$$
Therefore, the sum of the
 $F$-norm of the last two terms of (2.14) is not greater than
 $$ c\lambda\theta^2(1+{2\over
m})\biggl([[u_k]]_{(-\infty,-L]}+[[u_k]]_{[L,\infty)}\biggr).
$$
The obtained bounds imply that
 $u_{out,k}$ is a Cauchy sequence in the $F$-norm.
 Call the limit  $v.$ By passage to the limit we obtain
 $$
v(t) = u_{in}(t)-\lambda\int^{+\infty}_{-\infty}R(t-\tau)\ast
u^3(\tau) d\tau ,$$
whence $v(t)= u_{out}(t), $ as required.

Theorems 2.3 and 2.4 are proved.

\section{Riemann function estimates.
 \newline Proof of Lemmas 2.1 and 2.2}

Proof of Lemma 2.1.

$I(t,{\bf x})$ consists of an integral $I_S$ over the
surface of a light cone and an integral $I_C$ over the
interior of the cone.  Since $R(t,{\bf x})=-R(-t,{\bf
x})$, the integrals over forward and backward light cones
can be considered similarly.  The integral over the
surface of a light cone is the following  $$
I_S=\pm{1\over 4\pi}\int_{M_{1,\pm}}\int_{|{\bf x}-{\bf
y}|=|t-\tau|} u(\tau,{\bf y}) v(\tau,{\bf y}) w(\tau,{\bf
y}) dS{d\tau\over |t-\tau|} $$ and over the interior  $$
I_C=\pm\int_{M_{1,\pm}} \int_{|{\bf x}-{\bf y}|<|t-\tau|}
k(\mu) \,u(\tau,{\bf y}) v(\tau,{\bf y}) w(\tau,{\bf y})
d^3y d\tau, $$ where $k(\mu)=c\mu^{-1} J_1(m\mu),\;\;
J_1$ is the Bessel function, $\mu^2=(t-\tau)^2-|{\bf
x}-{\bf y}|^2$ and $M_{1,\pm}=M_1\cap \{\tau \,|\,
\pm(t-\tau)\geq 0 \}.$ The measure $dS$ is defined before
the formulation of  Lemma 2.1.

To the surface integral we apply Schwarz' inequality:
$$ I^2_S\leq\biggl(\int\int|w|^2 dS d\tau\biggr)
\biggl(\int\int |u|^2 |v|^2 (t-\tau)^{-2} dS
d\tau\biggr), $$ where the integrals are taken over the
range $$ |{\bf x}-{\bf y}|=|t-\tau|, \;\;\;\;\; \tau\in
M_1 $$ and $dS = \rho^2 d\omega, \rho\equiv|{\bf x}-{\bf
 y}|, |\omega|=1.$ The first factor on the right side is
bounded by  $$ {\langle w\rangle}^2_M = \sup_K \int_{
K\cap (M\times {I\!\!R}^3 ) } |w|^2 d{\tilde S}.$$ As for
the second factor, we first note that the integration by
part gives $$ \int_{|{\bf x}-{\bf y}|=|t-\tau|} \Phi({\bf
y}) dS = \int_{|{\bf x}-{\bf y}|=|t-\tau|} \Phi({\bf y})
|{\bf x}-{\bf y}| |t-\tau|^{-1}dS$$ $$=\int_{|{\bf
x}-{\bf y}|<|t-\tau|} {\partial\over\partial\rho}
(\Phi\rho^3)\, |t-\tau|^{-1} d\rho d\omega$$ $$ =
\int_{|{\bf x}-{\bf y}|<|t-\tau|} (\rho\Phi_{\rho} +
3\Phi) \,|t-\tau|^{-1}d^3y.  \eqno(3.1)$$

Applying this identity to  $\Phi=|v|^2=vv^*$ and using
$\rho\leq |t-\tau|$, we obtain $$\int_{|{\bf x}-{\bf
y}|=|t-\tau|} |v|^2 dS =\int_{|{\bf x}-{\bf y}|<|t-\tau|}
(\rho v v^*_{\rho} + \rho v_{\rho} v^* + 3v v^*)
|t-\tau|^{-1} d^3y $$ $$\leq \int_{|{\bf x}-{\bf
y}|<|t-\tau|} \;\bigl[\, 2|v v_{\rho}| + 3|v|^2\,
|t-\tau|^{-1}\, \bigr] d^3y$$ $$\leq
2\|v(\tau)\|_{2,*}\;\|v_{\rho}(\tau)\|_2 +3\delta^{-1}
\|v(\tau)\|_{2,*} .\eqno(3.2)$$ Therefore, $$ I^2_S\leq
c(\delta)\biggl(\int\int |u|^2 dS d\tau\biggr) \int_{M_1}
\|u(\tau)\|^2_\infty \, \|v(\tau)\|_e \|v(\tau)\|_{2,*}
\,(t-\tau)^{-2} d\tau.  $$ Here and in the following the
notation $\|\cdot\|_{p,*}$ means the $L_p$-norm of a
complex-valued function over the sphere $|{\bf x}-{\bf
y}|<|t-\tau|.$

Now consider the integral   $I_C$ over the interior. The
 contribution of the forward and backward cones is
estimated in the same way. Considering the light cone we
use the notation $$ \rho=|{\bf x}-{\bf y}|,\;
\mu^2=(t-\tau)^2-\rho^2,\; {\bf y}-{\bf x}=\rho\omega,$$
and introduce the light cone variables for the forward
and backward light cone $$ \xi=\pm(t-\tau)+\rho,\;\quad
\eta=\pm(t-\tau)-\rho,\;\quad \tau\in M_{1,\pm}.  $$ Thus
$\mu^2 = \xi \eta.$ We also introduce a weight factor:
  $$ l(\eta) = \cases{\eta^{3/2} \quad\mbox{ for }\quad
\eta\geq{1\over 2}\delta,\cr \eta^{3/4} \quad\mbox{ for
}\quad 0<\eta <{1\over 2}\delta.\cr} $$ Estimating the
contribution of the forward and backward cones by
Schwarz' inequality we obtain for each such contribution
 $|I_C|^2\leq AB,$ where \begin{eqnarray*} A&=&\int\int
{l(\eta)}^{-1} |w|^2 d\tau d^3y, \\
B&=&\int_{M_{1,\pm}}\int l(\eta) \,k(\mu)^2 \,|u|^2
\,|v|^2 \,d\tau d^3y.  \end{eqnarray*} Changing
variables, $(\rho,\theta)\to(\xi,\eta),$ we have
\begin{eqnarray*} A&=&{1\over 2}\int\int {l(\eta)}^{-1}
\int_{|\omega|=1} |w|^2\rho^2 d\omega d\xi d\eta\\
&\leq&{1\over 2}\int^{\infty}_0 \Bigl[\int_{K(t,{\bf
x},M_{\pm},\eta)} \int_{|\omega|=1} |w|^2\rho^2 d\omega
d\xi\Bigr] l(\eta)^{-1} d\eta .  \end{eqnarray*} The
integral in square brackets is precisely the integral of
$|w|^2$ over a part  $ K(t, {\bf x}, M_{\pm},\\ \eta)$ of
the surface of the forward or backward light cone
$K_{\pm}(t,{\bf x})$ with the top at  $(t,{\bf x}).$ This
part of the surface of the cone $K(t,{\bf
 x},M_{\pm,}\eta)$ is given by the condition $\tau =
\mp{(\xi+\eta)\over 2} + t \in M_{\pm}, $ $\eta \in
[0,+\infty).$ Since $l(\eta)^{-1}$ is integrable, the
expression for $A$ is bounded by  $$ A\leq c(\delta)
 \sup_K \int_{ K\cap (M\times {I\!\!R}^3 ) } |w|^2
 d{\tilde S},$$ where $K$ runs over all forward or
backward light cones and $d{\tilde S} $ denotes the usual
surface measure on the surface of $K.$ The factor
$c(\delta)$  appears due to the integral of
$l(\eta)^{-1}$, with depends on $\delta.$ In the second
factor $B,$ we use the gross asymptotic behavior of
 $k(\mu)=c\mu^{-1}J_1(m\mu)$ $$ k(\mu)^2\leq
c\mu^{-3}\leq c(\eta |t-\tau|)^{-3/2}, $$ see
\cite[Section 8.45]{22}. Therefore, $$ B\leq
c\int_{M_{1,\pm}} D(\tau) |t-\tau|^{-3/2} d\tau, $$ where
$$ D(\tau)=\int l(\eta) \eta^{-3/2} |u|^2 |v|^2 d^3y.  $$

We estimate  $D(\tau)$, dividing the domain for $\eta$ in
two parts:  $\eta>{1\over 2}\delta$ and $\eta<{1\over
2}\delta.$

The part of $D(\tau)$
over  $\eta>{1\over 2}$ is less than
$$ \mbox{ const }\|u(\tau)\|^2_{\infty}\;
\|v(\tau)\|^2_{2,*}.  $$

For  $\eta<{1\over 2}\delta,$ we have $l(\eta)
\eta^{-3/2} = \eta^{-3/4}.$ At each point we have the
identity $$\eta^{-3/4} |u|^2 |v|^2 = \mbox{div}_y
\biggl({{\bf x}-{\bf y}\over \rho} 4\eta^{1/4} u u^* v
v^* \biggr)$$ $$+ 4\eta^{1/4} \bigl((u u^*_{\rho} +
u_{\rho} u^*) |v|^2 + |u|^2 (v_{\rho} v^* + v
v^*_{\rho})\bigr) + 8\rho^{-1} \eta^{1/4} |u|^2 |v|^2.$$
Integration of this identity over the range
$0<\eta<{1\over 2}\delta$ (that is, over the spherical
shell  \-$|t-\tau|-{1\over 2}\delta < \rho < |t-\tau|)$
gives $$\int_{\eta<{1\over 2}\delta} \eta^{-3/4} |u|^2
|v|^2 d^3y =  4\int_{\eta={1\over 2}\delta} \delta^{1/4}
 |u|^2 |v|^2 dS$$ $$ +4\int_{\eta<{1\over 2}\delta}
(\eta^{1/4} (u u^*_{\rho} + u_{\rho} u^*) |v|^2 +
\eta^{1/4} |u|^2 (v_{\rho} v^* + v v^*_{\rho}) +
2\eta^{1/4} \rho^{-1} |u|^2 |v|^2) d^3y,\eqno(3.3)$$ the
contribution of the point $\eta=0$, that is, the
contribution of the spherical shell $\rho=|t-\tau|,$ is
 equal to zero.  To estimate the surface integral in
 (3.3) we use the identity (3.1) and, analogously to
(3.2),  we obtain the estimate
\begin{eqnarray*}\int_{\eta={1\over 2}\delta} |u|^2 |v|^2
dS &\leq& 2\|u(\tau)\|^2_{\infty} \|v(\tau)\|_e
\|v(\tau)\|_{2,*}\\ &&+(2+6\delta^{-1}) \|u(\tau)\|_e
\|v(\tau)\|_{2,*} \|u(\tau)\|_{\infty}\|v(\tau)\|_\infty.
\end{eqnarray*}

In the volume integral in (3.3) we use $\eta\leq{1\over
2}\delta$ and $\rho\geq\delta-\eta\geq{1\over 2}\delta.$
Therefore, we have  $$ D(\tau)\leq c(\delta)\biggl\{
\|u(\tau)\|^2_{\infty} \,\|v(\tau)\|_e
\,\|v(\tau)\|_{2,*} + \|u(\tau)\|_{\infty}
\,\|v(\tau)\|_{\infty} \,\|u(\tau)\|_e
\,\|v(\tau)\|_{2,*} \biggr\}.  $$

 Finally, we use the trivial estimate
 $$ \|v(\tau)\|_{2,*} \leq c\|v(\tau)\|_{\infty}
|t-\tau|^{3/2}, $$ which is raised to an arbitrary power
 $\alpha$. This is used to estimate the terms with
$\|\cdot\|_{2,*}$ both in the bound for
$D(\tau)$ and in the one for $I_S.$
Taking into account these estimates,
 we obtain the estimate of  Lemma 2.1.

Lemma 2.1 is proved.

\medskip
\noindent
Proof of  Lemma 2.2.

Denote by $W(t,{\bf x})$ the same integral as ${\cal
R}u(t,{\bf x})$ except that $u^3$ is to be replaced by $u
 v w$ and we shall obtain the estimates for this term.
 These estimates yields the estimates of Lemma.

According to the definition of the $F$-norm, $\|W\|_F$
consists of three terms (2.4).  To estimate the energy
norm, we apply the energy relation $$ \|R(t-\tau)*f\|_e =
\|f\|_2 $$ the function $f = u v w.$ We obtain
$$\|W(t)\|_e \leq \int_M \|u v w\|_2 d\tau \leq m^{-1}
\sup_{t\in M} \|w(\tau)\|_e \int_M \|u(\tau)\|_{\infty}
\|v(\tau)\|_{\infty} d\tau$$ $$\leq m^{-1}\sup_{\tau\in
M} \|w(\tau)\|_e \biggl(\int_M
\|u(\tau)\|^2_{\infty}d\tau\biggr)^{1/2}
\biggl(\int_M\|v(\tau)\|^2_{\infty} d\tau\biggr)^{1/2}.$$
We shall obtain the required estimates for Lemma 2.2(a)
by setting  $u=v=w$.  Using the relation $u^3-v^3 =
u^2(u-v)+ u v(u-v)+v^2(u-v)$ and taking instead of
$u,v,w,$ respectively, $u,u,v,$ or $u,v,u-v,$  or
$v,v,u-v,$ we obtain the desired estimates of Lemma
2.2(b) for this term.

To estimate the rest of  $F$-norm we write
$$
W = W_1+W_2,
$$
where
$$
W_1=\int_{M_1}\int R\, u\, v\, w\, d\tau d^3y,\;\quad\;
W_2=\int_{M_2}\int R\, u\, v\, w\, d\tau d^3y
$$
and
$$
M_1=[a,b]\cap\{\tau\;|\;|t-\tau|\geq 1\},\;\quad\;
M_2=[a,b]\cap\{\tau\;|\;|t-\tau|<1\}.
$$

To  $W_1$ we apply Lemma 2.1 with $\delta=1.$ Then $$
|W_1(t,x)| \leq  c{\langle w\rangle}_M \, \{\int_{M_1}
... d\tau \}^{1/2} $$ with the same integrand  as in
Lemma 2.1.  Since $\alpha < 1/6,\;$
$|t-\tau|^{-3/2+3\alpha}$ is integrable. Therefore
$$\|W_1(t)\|_{\infty} +
(\int^{+\infty}_{-\infty}\|W_1(t)\|^2_\infty dt)^{1/2}$$
$$\leq c{\langle w\rangle}_M \sup_{\tau\in
M}\|v(\tau)\|_e^{1/2-\alpha} [[v]]^\alpha_M
\Bigl\{[[u]]^2_M \, \sup_{\tau\in M} \|v(\tau)\|_e +
[[u]]_M [[v]]_M \sup_{\tau\in M} \|v(\tau)\|_e
\Bigr\}^{1/2} . $$ This implies Lemma 2.2(a) for $W_1$
when we set $u=v=w.$ Using the relation $u^3-v^3 =
u^2(u-v)+ u v(u-v)+v^2(u-v)$ and taking instead of
$u,v,w,$ respectively, $u,u-v,u,$ or $u,u-v,v,$  or
$v,u-v,v$ and setting $\alpha=0$ we obtain the desired
 estimates of Lemma 2.2(b).

Finally, let us estimate  $W_2(t,{\bf x}).$ We write it as
 $I_S+I_C$, where  $I_S$ is the integral over the surface of the
cone and $I_C$  is the integral over the interior of the cone.
 For  $I_S$ as in the proof of  Lemma 2.1, we use the integration
by parts $$ \int_{\rho=|t-\tau|}\Phi dS =
\int_{0\leq\rho\leq|t-\tau|}
{\partial\over\partial\rho}(\Phi\rho^2)d\rho d\omega =
\int_{0\leq\rho\leq|t-\tau|} (\Phi_{\rho}+2{\Phi\over\rho})d^3y
$$ and for $\Phi = u v w$ we have $$ I_S =
\int_{M_2}\int_{\rho\leq|t-\tau|}(u v w_{\rho} + u v_{\rho} w +
u_{\rho} v w + {2\over\rho}u v w) d^3y{d\tau\over|t-\tau|}\,
.\eqno(3.4) $$ Applying H\"older's inequality with exponents
$3,6,2$ or $3/2,6,6$ to the inner integral in (3.4) and using the
 estimates $$ \|u(\tau)\|_{3,*} |t-\tau|^{-1} \leq c
\|u(\tau)\|_{\infty} $$ and $$ \|\rho^{-1}u(\tau)\|_{{3\over
2},*} |t-\tau|^{-1} \leq c \|u(\tau)\|_{\infty} $$ for
$|t-\tau|\leq 1$, we obtain $$|I_S|\leq
c\int_{M_2}\bigl(\|u(\tau)\|_{\infty}\,\|v(\tau)\|_{6,*}
\,\|w_{\rho}(\tau)\|_{2,*} +
\|u(\tau)\|_{\infty}\,\|v_{\rho}(\tau)\|_{2,*}\,
\|w(\tau)\|_{6,*} $$ $$+\|v(\tau)\|_{\infty}
\,\|u_{\rho}(\tau)\|_{2,*}\,\|w(\tau)\|_{6,*} +
\|u(\tau)\|_{\infty}\,\|v(\tau)\|_{6,*} \,\|w(\tau)\|_{6,*}\bigr)
d\tau. $$ Taking into account that $J_1(m\mu) \mu^{-1} = {\it
O}(\mu^{-3/2})$ (see \cite[Section 8.45]{22} ) we have for the
integral over the interior of the cone $$ |I_C| \leq
c\int_{M_2}\|u(\tau)\|_{\infty}
\,\|v(\tau)\|_{2,*}\,\|w(\tau)\|_{2,*} \,d\tau.  $$ As in the
proof of Lemma 2.1 the asterisks indicate the integral in the
norm over $\rho\leq |t-\tau|$ only.  We take into account that
for $ |t-\tau|\leq 1\;$\quad $\|u(\tau)\|_{p,*}\leq
c\|u(\tau)\|_{\infty} ,$ and that the integration is taken  over
$M_2 = [a,b]\cap\{\tau\;|\;|t-\tau|<1\}$ only.  Then, we set
 $u=v=w$ and obtain $$|W_2(t,{\bf x})| \leq
c_1\biggl(\int_{M_2}\|u(\tau)\|^2_{\infty} d\tau \biggr)
\sup_{\tau\in[a,b]} \|u(\tau)\|_e$$ $$\leq c_2\biggl(\int_{M_2}
\|u(\tau)\|^2_{\infty} d\tau\biggr)^{1/2} \sup_{\tau\in M}
\|u(\tau)\|_{\infty}\; \sup_{\tau\in M} \|u(\tau)\|_e.  $$ Making
in the integral over $t,\tau$ the change of variables on
$t-\tau,\tau$, we obtain $$\biggl(\int^{+\infty}_{
-\infty}\sup_{{\bf x}}|W_2(t,{\bf x})|^2dt\biggr)^{1/2} \leq
c\biggl(\int_M\|u(\tau)\|^2_{\infty} d\tau\biggr)^{1/2}
\sup_{\tau\in M}\|u(\tau)\|_{\infty}\sup_{\tau\in M}\|u(\tau)\|_e
.$$ $$\leq c[[u]]^2_M \|u\|_{F,M} .  $$ This yields the part (a)
of Lemma.

On the other hand, for the part (b) we use Sobolev's
inequality,
 $\|u\|_{6,*}\leq c\|u\|_e$ and the relation
$$\int_{M_2}\|u(\tau)\|_{\infty}
d\tau\leq\biggl(\int_{M_2}\|u(\tau)\|^2_{\infty}
d\tau\biggr)^{1/2}\biggl(\int_{M_2}d\tau\biggr)^{1/2}.$$
Thus we get  $$|W_2(t,{\bf x})| \leq
c_1\biggl(\Bigl(\int_{M_2}\|u(\tau)\|_{\infty}d\tau\Bigr)
\sup_{\tau\in M}\|v(\tau)\|_e \,\sup_{\tau\in
M}\|w(\tau)\|_e$$ $$+\Bigl(\int_{M_2}\|v(\tau)\|_{\infty}
d\tau \Bigr) \sup_{\tau\in M}\|u(\tau)\|_e
\,\sup_{\tau\in M}\|w(\tau)\|_e\biggr)$$ $$\leq
c_2\biggl(\Bigl(\int_{M_2}\|u(\tau)\|_{\infty}
d\tau\Bigr)^{1/2} \sup_{\tau\in M}\|v(\tau)\|_e
\,\sup_{\tau\in M}\|w(\tau)\|_e$$
$$+\Bigl(\int_{M_2}\|v(\tau)\|_{\infty} d\tau\Bigr)^{1/2}
\sup_{\tau\in M}\|u(\tau)\|_e \,\sup_{\tau\in
M}\|w(\tau)\|_e\biggr),$$ whence $$
\int^{+\infty}_{-\infty}|W_2(t,{\bf x})|^2 dt \leq
c\Bigl( [[u]]_M \,\sup_{\tau\in M}\|v(\tau\|_e
\,\sup_{\tau\in M}\|w(\tau)\|_e$$ $$+ [[v]]_M
\sup_{\tau\in M}\|u(\tau)\|_e \,\sup_{\tau \in
M}\|w(\tau)\|_e \Bigr).$$ Again using the relation
$u^3-v^3 = u^2(u-v)+ u v(u-v)+v^2(u-v)$ and taking
instead of $u,v,w,$ respectively, $u,u,u-v,$ or
$u,v,u-v,$  or $v,v,u-v,$ we obtain the estimate of Lemma
 2.2(b).

Lemma 2.2 is proved.

\section{
 Construction of
the       quantum field \newline as a bilinear form}

To construct the quantum field as a bilinear form we
shall start from the quantum non-linear wave equation
written in the from of integral equation (1.2).

We begin with a brief sketch and an outline of the construction
of solution of Equation (1.2) and then we turn to the description
 of the technical details.

We shall construct the solution $\phi(t,{\bf x})$  of
 Equation (1.2) as a bilinear form. This bilinear form is
defined in the Fock space ${\cal H}_{in}$ of the free
field $\phi_{in}$ and can be expanded in terms of
creation and annihilation operators.  By $:\;\; :$ in
(1.2) we denote the normal ordering with  respect to the
free field $\phi_{in},$ and correspondingly, by product
we mean the normal ordered product of the bilinear forms.
However, an operator-valued structure of the interacting
field is unknown in advance.

Since, in fact, we come from the notion of wave operator, so the
natural initial quantum field should be the free quantum
$in$-field that enters  into Equation (1.2)).

Thus, we need to construct the bilinear form that
corresponded to the interacting quantum field and is
defined on the whole space-time, that is, to construct
the unique solution in the large, that corresponds to the
unique initial $in$-field.

A representation of the solution in the form of a limit
of some iterative series is a natural way of the
construction of this solution. We construct the iterative
series as series expanded in terms of Wick polynomials on
the free $in$-field. Therefore, to obtain the solution in
the large it is sufficient to construct a bilinear form
corresponding to the interacting field at any time. It
can be continued  in the large by translation with the
Hamiltonian.  Nevertheless, we would like to obtain the
solution in the large defined as a bilinear  form. It is
convenient to approximate the solution by bilinear forms
defined for all times.

It turns out that it is possible. This is connected with
the fact that coherent vectors are the eigenvectors of
annihilation operators.  Moreover, the Wick polynomial of
the free  field is a bilinear form, the free field is the
sum of the creation and annihilation operators, and the
creation operator is conjugate to the annihilation
operator. Taking this into account, we obtain that the
matrix elements between coherent vectors are equal to the
corresponding polynomial depending on the sum of
corresponding eigenvalue of one coherent vector and the
complex conjugate of the eigenvalue of another coherent
 vector.

Therefore, if we consider iterations of the right side of
Equation (1.2), that is, the expressions $$
\phi^{(l)}(t,{\bf x}) = \{\phi_{in}+\lambda
N_R\{\phi_{in}+\lambda N_R\{...\{\phi_{in}+\lambda
N_R(\phi_{in})\}...\}\}\}_{(l)}(t,{\bf x})\eqno(4.1) $$
where $$ N_R(\phi)=-\int^t_{-\infty}\int R(t-\tau,{\bf
x}-{\bf y}):\!\phi^3(\tau,{\bf y})\!:d\tau d^3y, $$ we
approximate the quantum field by Wick polynomials, i.e.
by bilinear forms.  These bilinear forms
$\phi^{(l)}(t,{\bf x})$ are defined on some  sufficiently
wide dense subspaces, in particular, on the subspace
generated by linear combination of coherent vectors.
Consider  matrix elements of the constructing bilinear
form on the vectors that are equal to a finite linear
combination of coherent vectors near to the vacuum we
reduce, in fact,  these matrix elements to a bilinear
combination of iterations. These iterations are the
iterations of corresponding solution of classical wave
equation with small complex $in$-data. This allows us to
use the theorems proved in Sect. 2.

To prove the convergence of approximations we choose as
convenient subspace the set of linear combinations of
coherent vectors near to the vacuum (we denote it by
 $D_\theta$).  This subspace is dense in the Fock space.
We  define explicitly the quantum field on this subspace
and with the help of weak estimates of Sect. 2 and 3 we
prove the convergence of the approximations (4.1).  It is
convenient to introduce            in addition
 approximations with  space-time and an ultraviolet
cut-offs.

Bilinear forms generated by creation and annihilation
operators was considered by Kristensen, Mejlbo and
Poulsen \cite{1}-\cite{3}.  Baez in \cite{10} stated and
 proved the results that we need about Wick polynomials
 as bilinear forms. These results can be applied to the
approximations that we consider.  Note that we  use
slightly other notations  as Baez \cite{10}.

Therefore, we construct the  quantum field $\phi(t,{\bf
 x})$ as a bilinear form on $D_\theta\times D_\theta$ and
 approximate it by  bilinear form corresponding to the
iterations (4.1) (with an ultraviolet and space-time
cut-off). The limit of these iterations and cut-offs
converges and gives the bilinear form, that is, the
solution of (1.1) and (1.2).

Let pass to the detailed presentation. Introduce the
notations that we need.  Let  ${\cal H}_{in}$ be the Fock
Hilbert space of the free $in$-field.  The field
 $\phi_{in}(t,{\bf x})$ in terms of the annihilation  $a$
and the creation $a^+$ operator has the following form
(we shall use the notation of \cite[гл. Х, \S{7}]{19} and
shall not write in the following the  index ``$in$'' for
the creation and annihilation operators):  $$
\phi_{in}(t,{\bf
x})={1\over(2\pi)^{3/2}}\int(e^{-i\mu({\bf p})t+i{\bf
p}{\bf x}}a({\bf p}) + e^{+i\mu({\bf p})t-i{\bf p}{\bf
x}}a^+({\bf p})) {d^3 p\over\sqrt{2\mu ({\bf p})}}\;, $$
where $\mu({\bf p})=({\bf p}^2+m^2)^{1/2},$ $$[a({\bf
p}),\, a^+({\bf p'})]= \delta ({\bf p}-{\bf p'})$$ $$
[a({\bf p}), \,a({\bf p'})] = [a^+({\bf p}),\,a^+({\bf
p'})]=0.  $$ In $(t,{\bf x})$-space it is convenient to
introduce also the notation for positive- and negative
parts $$\phi^+_{in}(t,{\bf x}) = A^+(t,{\bf x}) = {1\over
(2\pi)^{3/2}}\int e^{i\mu({\bf p})t-i{\bf p}{\bf
x}}a^+({\bf p}){d^3p\over \sqrt{2\mu({\bf p})}} , $$ $$
\phi^-_{in}(t,{\bf x}) = A(t,{\bf x}) = {1\over
(2\pi)^{3/2}}\int e^{-i\mu({\bf p})t+i{\bf p}{\bf
x}}a({\bf p}){d^3p\over \sqrt{2\mu({\bf p})}}.  $$

Let a pair of complex $z_1,z_2\in {\cal S}({I\!\!R}^3)$
	is given, denote by $$u_{in}(t,{\bf x},z_1,z_2) =
 (2\pi)^{-3/2}\int\Bigl(e^{-i\mu({\bf p})+i{\bf p}{\bf
x}} z_1\tilde{ }\;({\bf p}) + e^{+i\mu({\bf p})-i{\bf
p}{\bf x}} z_2\tilde{ }\;({\bf p})\Bigr) {d^3{\bf
p}\over\sqrt{2\mu({\bf p})}} $$ the complex solution
corresponding to a pair $(z_1,z_2).$ This solution, or
its initial data, defines uniquely a pair $(z_1,z_2)$,
which corresponds to the the positive- and negative parts
 of $u_{in}(t,{\bf x},z_1,z_2),$ $$
z_1(\cdot)=2^{-1/2}(2\pi)^{3/2}(-\Delta + m^2)^{1/4}
u_{in}(0,-(\cdot)) + i 2^{-1/2}(2\pi)^{3/2}(-\Delta +
 m^2)^{-1/4}{\dot u}_{in}(0,-(\cdot)) $$ $$
z_2(\cdot)=2^{-1/2}(2\pi)^{3/2}(-\Delta + m^2)^{1/4}
u_{in}(0,\cdot) - i 2^{-1/2}(2\pi)^{3/2}(-\Delta +
 m^2)^{-1/4}{\dot u}_{in}(0,\cdot).  $$

Let
 $u(t,{\bf x},z_1,z_2)$
denotes the solution of (2.3) and
$u_{out}(t,{\bf x},z_1,z_2)$ denotes the $out$-data
 corresponding to the initial $in$-data
  $u_{in}(t,{\bf x},z_1,z_2)$, or,
that is equivalent, corresponding to the pair
  $(z_1,z_2).$

To define the bilinear forms
 we introduce the
convenient dense subspaces in
 the Fock Hilbert space
 ${\cal H}_{in}$ of the free $in$-field.
Define first of all the coherent vectors.
Let
$$ |z\rangle = \exp{(z a^+)}\Omega,\;\quad\quad
\Omega=|0\rangle = \; vacuum .\eqno(4.2) $$ Here $z
a^+=\int z \tilde{ }\;(k) a^+(k) d^3 k,\;\; z\in{\cal
S},$ where ${\cal S}$ is the Schwartz space of rapidly
decreasing smooth complex-valued functions on
 ${I\!\!R}^3$, and a tilde denotes the Fourier transform.
Equation (4.2) implies that the scalar product in ${\cal
 H}_{in}$ of two coherent vectors is equal to $$ \langle
z_1 |z_2\rangle =  \exp({\overline z}_1, z_2) =
\exp(\int{\overline z}_1({\bf x}) z_2({\bf x})d^3 x).  $$
Let $D$ be the subspace of all finite linear combinations
of coherent vectors $$ D=\Bigl\{\chi\in{\cal H}_{in}\;
 |\; \chi = \sum\alpha_j |z_j\rangle ,\;\; z_j\in{\cal
S}({I\!\!R}^3)\Bigr\}.  $$ We introduce also the subspace
 $D_\theta$, of linear combinations of coherent  vectors
 near to the vacuum, \begin{eqnarray*} D_\theta =
 \Bigl\{\chi\in D &\;|\;& \chi = \sum\alpha_j
|z_j\rangle, z_j\in {\cal S}({I\!\!R}^3),\cr &
&\|u_{in}(\cdot,\cdot,z_j,0)\|_F < \theta/4,
\;\quad\;\|u_{in}(\cdot,\cdot,0,z_j)\|_F <
\theta/4\Bigl\}.  \end{eqnarray*}

It is clear that the subspace
 $D_\theta$ is dense in  ${\cal H}_{in}$.

It is implied, for instance, by the fact  that
derivatives of coherent vectors are finite-particle
vectors, $${d^n\over d\alpha^n}|\alpha z\rangle
\biggm|_{\alpha=0} = (z a^+)^n\Omega,$$ and any vector
 ${\cal H}_{in}$ can be approximated by finite-particle
vectors (from the Schwartz space).  Derivatives
themselves can be approximated by linear combination of
 coherent vectors near to  vacuum. Moreover, for all
positive integer $n \;\;$ $D_{\theta}\subset
D(H^n_{in})$, where  $H_{in}$ is the free Hamiltonian of
the $in$-field. In addition, for all positive $n\,\,$
 $H^n_{in}$ is essentially self-adjoint on $D_\theta.$
The inclusion $D_\theta\subset D(H^n_{in})$ is the
consequence of the simple bound \begin{eqnarray*}
\||H^n_{in}|z\rangle\|^2 & =
&\sum_k\|\Bigl(\sum_i\mu(p_i)\Bigr)^n {{z\tilde{
}\;(p_1)...z\tilde{ }\;(p_k)}\over {k!^{1/2}}}\|^2 \\ &
\leq &\sum_k{1\over k!}\Bigl(\sum{n!\over
n_1!...n_k!}\Bigl(\sup_{0\leq j\leq n}\|\mu(p)^j z\tilde{
}\;(p)\|\Bigr)^k\Bigr)^2\\ &\leq& \sum_k {k^{2n}\over
k!}\sup_{0\leq j\leq n}\|(-\Delta+m^2)^{j/2} z\|^{2 k}  <
\infty.  \end{eqnarray*} The self-adjointness of
$H^n_{in}$ on  $D_{\theta}$ follows from the Nelson's
theorem. A dense subspace of analytical vectors for
 $H^n_{in}$ is the space of linear combinations of
coherent vectors near to the vacuum with test functions
with compact support in the momentum space.

\medskip
Thus, we formulate the main theorem.

\medskip
Theorem 4.1.

Let $$ \chi_1=\sum^{n_1}_{j=1}\alpha_j e^{(v_j
a^+)}\Omega,\; \quad \chi_2=\sum^{n_2}_{k=1}\beta_k
e^{(w_k a^+)}\Omega,\eqno(4.3) $$ where complex-valued
functions $v_j,w_k\in{\cal S}({I\!\!R}^3)$ and for the
constant $\theta$  from Theorems 2.3 and 2.4
$$\|u_{in}(\cdot,\cdot,v_j,0)\|_F < {\theta \over 4}\, ,
\quad \|u_{in}(\cdot,\cdot,0,v_j) \|_F < {\theta \over
4}\, ,      $$ $$ \|u_{in}(\cdot,\cdot,w_k,0)\|_F <
{\theta \over 4}\, ,           \quad
\|u_{in}(\cdot,\cdot,0,w_k)\|_F < {\theta \over 4}\, ,
\eqno(4.4) $$ for all $j,k.$

Then the following expressions give  bilinear forms $$
\phi(t,{\bf x})(\chi_1,\chi_2) =
\sum_{j,k}\overline{\alpha}_j \beta_k
e^{(\overline{v}_j,w_k)} u(t,{\bf x},{\bar
v}_j,w_k),\eqno(4.5) $$ $$ :\!\phi^3(t,{\bf
x})\!:(\chi_1,\chi_2) = \sum_{j,k}\overline{\alpha}_j
\beta_k e^{({\bar v}_j,w_k)}u^3(t,{\bf x},{\bar
v}_j,w_k),\eqno(4.6) $$ $$ \phi_{out}(t,{\bf
x})(\chi_1,\chi_2) = \sum_{j,k}\overline{\alpha}_j
\beta_k e^{(\overline{v}_j,w_k)} u_{out}(t,{\bf x},{\bar
v}_j,w_k).\eqno(4.7) $$ These bilinear form are
symmetrical and are defined  on $D_\theta\times
 D_\theta$. Moreover, these bilinear forms satisfies the
following equalities $$ \phi(t,{\bf x}) =
\phi_{in}(t,{\bf x})-\lambda\int^t_{-\infty}\int
R(t-\tau,{\bf x}-{\bf y}):\!\phi^3(\tau,{\bf y})\!:d\tau
d^3y, \eqno(4.8) $$ $$\phi_{out}(t,{\bf x}) =\phi(t,{\bf
x})-\lambda\int^\infty_t R(t-\tau,{\bf x}-{\bf
y}):\!\phi^3(\tau,{\bf y})\!:d\tau d^3y $$ $$=
\phi_{in}(t,{\bf x})-\lambda\int^\infty_{-\infty}\int
R(t-\tau,{\bf x}-{\bf y}):\!\phi^3(\tau,{\bf y})\!:d\tau
d^3y $$ $$= \phi_{in}(t,{\bf x})+\int^{+\infty}_{-\infty}
R(t-\tau,{\bf x}-{\bf y})(\Box+m^2)\phi(\tau,{\bf
y})d\tau d^3y, \eqno(4.9)  $$ and  $\phi_{out}(t,{\bf
x})$ satisfies the free equation $$
(\Box+m^2)\phi_{out}(t,{\bf x}) = 0.  $$ In addition, on
 $D_\theta\times D_\theta$ the bilinear form
 $\phi(t+T,{\bf x})$ converges to the  bilinear form
$\phi_{out}(t,{\bf x})$ as $T\to +\infty$, $$
\phi(t+T,{\bf x})(\chi_1,\chi_2)\stackrel
{\|\cdot\|_{e}}{\rightarrow}\phi_{out}(t,{\bf
x})(\chi_1,\chi_2) \quad\mbox{ при }\quad T\to
+\infty.\eqno(4.10) $$

\medskip
Proof of Theorem 4.1.

To prove that (4.5)-(4.7) define a bilinear form we use
the approximations (4.1), spatial, time, and ultraviolet
cut-offs.  Then we obtain an approximation of bilinear
form, that is given by the solutions corresponding to the
smooth initial $in$-data with compact support, i.e. by
solutions that are analogous to the solutions that appear
in the proof of Theorem 2.3, 2.4 with initial $in$-data
that belong to ${\cal F}^C.$

This approximation may be obtained in the following.

We change the integral over $(-\infty,t]$ on the integral
 over $[S,t]$.  This change corresponds to the time
cut-off.  Let $\phi_{in,\sigma,\Lambda}(t,{\bf x})
=\phi_{in,\sigma}(t,{\bf x})\Lambda({\bf x})$, where
$\phi_{in,\sigma}(t,{\bf x}) = \int\phi_{in}(t,{\bf
x}-{\bf y})\sigma ({\bf y}) d^3y$, here reals functions
$\sigma,\Lambda\in{\cal S}({I\!\!R}^3)$ and $\Lambda({\bf
x})$ has a compact support.  The change of
 $\phi_{in}(t,{\bf x})$ on
 $\phi_{in,\sigma,\Lambda}(t,{\bf x})$ corresponds to an
 ultraviolet and volume cut-offs.  These changes
correspond to the approximation by  bilinear forms given
by solutions with smooth initial $in$-data with compact
support.  Finally, we approximate the field solution by
bilinear forms $\phi^{(l)}_{S,\sigma,\Lambda} (t,{\bf
 x})$, where $$ \phi^{(l)}_{S,\sigma,\Lambda}(t,{\bf x})
=\Bigl\{\phi_{in,\sigma,\Lambda}+\lambda
N_{R,S}\Bigl\{...\Bigl\{\phi_{in,\sigma,\Lambda}+\lambda
N_{R,S}(\phi_{in,\sigma,\Lambda})\Bigr\}...
\Bigr\}\Bigr\}_{(l)}(t,{\bf x})\eqno(4.11) $$ and $$
N_{R,S}(\phi)=-\int^t_S\int R(t-\tau,{\bf x}-{\bf
y}):\!\phi^3(\tau,{\bf y})\!:d\tau d^3y.  $$ (4.11)
contains Wick polynomial of $\phi_{in,\sigma,\Lambda}.$
It is clear that these Wick polynomials are correctly
defined bilinear forms, see, for instance, \cite{4},
\cite[Theorem 3]{10}, \cite[ch. Х, \S{7}]{19}, \cite{23}.

Let us write how
 these bilinear forms  generated by Wick polynomials
act in the Fock space.
These bilinear forms have the following Wick symbols.
Let $\chi_1, \chi_2\in D^\infty(H_{in})$, then for the
bilinear form $:\!\phi_{in}(t_1,{\bf
x}_1)...\phi_{in}(t_n,{\bf x}_n)\!:$ we have $$
:\!\phi_{in}(t_1,{\bf x}_1) \dots \phi_{in}(t_n,{\bf
x}_n)\!:(\chi_1,\chi_2) $$ $$ =
\sum_{I\subset\{1,...,n\}} \Bigl(\prod_{i\in I}
A(t_i,{\bf x}_i)\;\chi_1, \prod_{i\in\{1,...,n\}\setminus
I} A(t_i,{\bf x}_i)\;\chi_2\Bigr)\, ,\eqno(4.15) $$ (see
[10, Theorem 3], [24, ch. X, \S{7}]).

It is easy to see that $$ u_{in}(t,{\bf x},z_1+z_2) =
u_{in}(t,{\bf x},z_1,0)+u_{in}(t,{\bf x},0,z_2), $$ $$
\overline{u_{in}(t,{\bf x},z_1,z_2)} = u_{in}(t,{\bf
x},\overline z_2,\overline z_1)\eqno(4.13) $$ и $$
A(t,{\bf x})e^{(z a^+)}\Omega=u_{in}(t,{\bf x},0,z)e^{(z
a^+)}\Omega.  $$ The last relation is implied by the
following simple calculation \begin{eqnarray*} a(k)e^{(z
a^+)}\Omega &=& a(k)\sum^\infty_{n=0}{(z a^+)^n\over
n!}\Omega=\sum^\infty_{n=0}\Bigl[a(k),{(z a^+)^n\over
n!}\Bigr]\Omega\\ &=&\sum^\infty_{n=1} z\tilde{ }\;(k)
{(z a^+)^{n-1}\over (n-1)!}\Omega = z\tilde{ }\;(k)e^{(z
a^+)}\Omega, \end{eqnarray*}  see, for instance,
\cite[ch. 9.1]{23}.

Therefore, if $\chi_1=\sum_j\alpha_j|v_j\rangle,\;
\chi_2=\sum_k\beta_k|w_k\rangle,$ then  (4.12) gives
$$
:\!\phi_{in}(t_1,{\bf x}_1)\dots \phi_{in}(t_n,{\bf
x}_n)\!: (\chi_1,\chi_2)  $$
$$
= \sum_{j,k}{\overline\alpha}_j\beta_k
\sum_{I\subset\{1,...,n\}}
\Bigl(\prod_{i\in
I} u_{in} (t_i,{\bf x}_i,0,v_j)|v_j\rangle,
\prod_{i\in \{1,...,n\}\backslash I}
u_{in} (t_i,{\bf x}_i,0,w_j)|w_j\rangle\Bigr) $$
$$ = \sum_{j,k}{\overline\alpha}_j\beta_k\langle
v_j|w_k\rangle \prod_{i\in\{1,...,n\}}
\Bigl(\,\overline{u_{in} (t_i,{\bf x}_i,0,v_j)} +
u_{in}(t_i,{\bf x}_i,0,w_k)\Bigr) $$ $$ =
\sum_{j,k}{\overline \alpha}_j\beta_k\langle
v_j|w_k\rangle \prod_{i\in\{1,...,n\}} u_{in}(t_i,{\bf
x}_i,{\overline v}_j,w_k).\eqno(4.14) $$ In particular,
it follows from  (4.14) that $$
:\!\phi_{in,\sigma,\Lambda}(t_1,{\bf
x}_1)\dots\phi_{in,\sigma,\Lambda}(t_n,{\bf
x}_n)\!:(\chi_1,\chi_2) $$ $$= \sum_{j,k}{\overline
\alpha}_j\beta_k\langle v_j|w_k\rangle
\prod_{i\in\{1,...,n\}} u_{in,\sigma,\Lambda} (t_i,{\bf
x}_i,{\overline v}_j,w_k),\eqno(4.15) $$ where
$u_{in,\sigma,\Lambda}(t,{\bf x}$ is a free solution with
the Cauchy data at time zero given by
 $$u_{in,\sigma,\Lambda}(0,{\bf x},\overline{v},w) =
\Lambda({\bf x}) \int u_{in}(0,{\bf x}-{\bf
y},\overline{v},w)\,\sigma({\bf y}) \,d^3y,$$ $${\dot
 u}_{in,\sigma,\Lambda}(0,{\bf x},\overline{v},w) =
\Lambda({\bf x}) \int {\dot u}_{in}(0,{\bf x}-{\bf
y},\overline{v},w)\,\sigma({\bf y}) \,d^3y,$$

The relations (4.14), (4.15) imply that on $D\times D$
the bilinear forms-iterations (4.11)  satisfy the
 equality $$ \phi^{(l)}_{S,\sigma,\Lambda}(t,{\bf
x})(\chi_1,\chi_2) = \sum_{j,k}{\overline
\alpha}_j\beta_k\langle v_j|w_k\rangle
\,u^{(l)}_{S,\sigma, \Lambda}(t,{\bf
x},\overline{v}_j,w_k), $$ where
$u^{(l)}_{S,\sigma,\Lambda}(t,{\bf
x},\overline{v}_j,w_k)$ is the $l$th-iteration of
Equation (2.1) - (2.2)  with the Cauchy data at time $S$
equal to  the Cauchy data of the free solution
$u_{in,\sigma,\Lambda}(t,{\bf x},\overline{v}_j,w_k)$.

Now we show that for
$l\to \infty$ the bilinear forms
$\phi^{(l)}_{S,\sigma,\Lambda}$ and
$:\!\phi^{(l)3}_{S,\sigma,\Lambda}(t,{\bf x})\!:$
converge
 on $D_\theta\times D_\theta$
to the bilinear forms
$\phi_{S,\sigma,\Lambda}(t,{\bf x})$ and $
:\!\phi^{3}_{S,\sigma,\Lambda}(t,{\bf x})\!:\;$
and  on vectors of the form (4.3)
 these bilinear forms are equal to
$$
\phi_{S,\sigma,\Lambda}(t,{\bf x})(\chi_1,\chi_2)
=\sum_{j,k}{\overline \alpha}_j\beta_k
e^{({\overline v}_j,w_k)}
u_{S,\sigma,\Lambda}(t,{\bf x},\overline{v}_j,w_k),
\eqno(4.16)$$ $$ :\!\phi^{3}_{S,\sigma,\Lambda}(t,{\bf
x})\!:(\chi_1,\chi_2) = \sum_{j,k}{\overline \alpha}_j\beta_k
e^{({\overline v}_j,w_k)}u^{3}_{S,\sigma,\Lambda}(t,{\bf
x},{\overline v}_j,w_k), \eqno(4.17) $$  where
$u_{S,\sigma,\Lambda}(t,{\bf x},{\bar v}_j,w_k)$ is the
solution of (2.1) - (2.2) with the Cauchy data at time
$S$ equal to the Cauchy data  of the free solution
 $u_{in,\sigma, \Lambda}(t,{\bf x},{\bar v}_j,w_k).$ In
 other words, these bilinear forms satisfy the following
equation $$ \phi_{S,\sigma,\Lambda}(t,{\bf x})
=\phi_{in,\sigma,\Lambda}(t,{\bf x})-\lambda\int^t_S\int
R(t-\tau,{\bf x}-{\bf y}):\!\phi^3_{S,\sigma,\Lambda}
(\tau,{\bf y})\!: d\tau d^3y.  $$ Really, first we note
that $D_\theta \subset D \subset D^\infty(H_{in})$ and,
thus, the approximations $\phi^{(l)}_{S,\sigma,\Lambda}$
are correctly defined on $D_\theta \times D_\theta.$
Choose, then, an ultraviolet  and space cut-offs such,
that for the states from $D_\theta \times D_\theta$  the
$F$-norm of $u_{in,\sigma,\Lambda}(t,{\bf x},{\bar
v}_j,w_k)$ would be less than  $\theta/2.$ For this
 purpose,   we shall take $\Lambda$ with compact support
 and $\sigma$ such that $\|\sigma\|_1\leq 1,$\quad
 $\|\Lambda\|_\infty\leq 1. $ It is obvious that $$
\|u_{in,\sigma,\Lambda}(\cdot,\cdot,\overline{ v}_j,w_k)
- u_{in}(\cdot,\cdot,\overline{v}_j,w_k)\|_F \to
0\eqno(4.18) $$ for such $\Lambda$  and $\sigma$ and for
$\Lambda \to 1,$ $\sigma \to$ $\delta$-function in ${\cal
S}'$ (our choice is  $v_j,w_k \in {\cal S}({I\!\!R}^3)$).

Therefore, the inequalities (4.4) for
$u_{in,\sigma,\Lambda}(\cdot,\cdot,\overline{v}_j,w_k)$
are fulfilled for the states  from $D_{\theta}\times
D_{\theta}$ and with the change $\theta$ on $2\theta$,
and, thus, the conditions of Theorem (2.4) are fulfilled
and the approximations $u^{(l)}_{S,\sigma,\Lambda}(t,{\bf
x},\overline{v}_j,w_k)$ correspond to the initial data at
time $S$ equal to
$u_{in,\sigma,\Lambda}(\cdot,\cdot,\overline{v}_j,w_k)$.
These initial data are smooth and have  compact support.
As in the proof of Theorem 2.4 we obtain,  that the
approximations $u^{(l)}_{S,\sigma,\Lambda}(t,{\bf
x},\overline{v}_j,w_k)$ converge to
$u_{S,\sigma,\Lambda}(t,{\bf x},\overline{v}_j,w_k)$ for
$l \to \infty.$ This means that the  bilinear forms
$\phi^{(l)}_{S,\sigma,\Lambda}(t,{\bf x})$ and
$:\!\phi^{(l)3}_{S,\sigma,\Lambda}(t,{\bf x})\!:$
converge to the bilinear forms (4.16)-(4.17).

Now let $S \to -\infty,$
we obtain, as in the proof of Theorem 2.4,
that as  $S \to -\infty \;$ \quad
$ u_{S,\sigma,\Lambda}(t,{\bf x},\overline{v}_j,w_k)$
converges to the solution
$u_{\sigma,\Lambda}(t,{\bf x},\overline{v}_j,w_k) $
with $in$-data
$u_{in,\sigma,\Lambda}(\cdot,\cdot,\overline{v}_j,w_k).$
On $D_\theta \times D_\theta$ this gives the convergence of
the bilinear forms $\phi_{S,\sigma,\Lambda}(t,{\bf x})$
and $:\!\phi^3_{S,\sigma,\Lambda}(t,{\bf x})\!:$
to the bilinear forms $\phi_{\sigma,\Lambda}(t,{\bf x}),$
\quad $:\!\phi^3_{\sigma,\Lambda}(t,{\bf x})\!:.$

On vectors (4.3) from $D_\theta$ these bilinear forms are
equal to $$ \phi_{\sigma,\Lambda}(t,{\bf
x})(\chi_1,\chi_2)=\sum_{j,k}\overline{ \alpha}\beta_k
e^{(\overline{v}_j,w_k)}u_{\sigma,\Lambda}(t,{\bf
x},\overline{v}_j,w_k), $$ $$ :\!\phi^3_{\sigma,
\Lambda}(t,{\bf
x})\!:(\chi_1,\chi_2)=\sum_{j,k}{\overline \alpha}\beta_k
e^{({\overline v}_j,w_k)}u^{3}_{\sigma,\Lambda}(t,{\bf
x},\overline{v}_j,w_k), $$ where
 $u_{\sigma,\Lambda}(t,{\bf x},\overline{v}_j,w_k)$ is
the solution of (2.1) - (2.2) with $in$-data
$u_{in,\sigma,\Lambda}(t,{\bf x},\overline{v}_j,w_k)$,
and satisfy the equation $$ \phi_{\sigma,\Lambda}(t,{\bf
 x}) =\phi_{in,\sigma,\Lambda}(t,{\bf x})
-\lambda\int^t_{-\infty}\int R(t-\tau,{\bf x}-{\bf
y}):\!\phi^3(\tau,{\bf y})\!:d\tau d^3y.  $$ Finally let
$\Lambda$ tend to $1$ and $ \sigma$ to $\delta$-function.
Since (4.18) fulfills, so
$u_{\sigma,\Lambda}(\cdot,\cdot,\overline{v}_j,w_k)\to
u(\cdot,\cdot,\overline{v}_j,w_k)$ in $F$-norm as
 $\Lambda\to 1,$~ $\sigma\to \delta$-function. But this
means that  $\phi_{\sigma,\Lambda}(t,{\bf x})$ and
$:\!\phi^3_{\sigma,\Lambda}(t,{\bf x})\!:$ converge on
 $D_\theta\times D_\theta$ to $\phi(t,{\bf x})$ and
$:\!\phi^3(t,{\bf x})\!:$.

Now we construct the bilinear form for $\phi_{out}(t,{\bf
x}).$ Equation (4.9) defines $\phi_{out}(t,{\bf x})$ as a
 bilinear form and (4.5)-(4.6) and Theorem 2.4 imply that
the bilinear form is given by the equality (4.7) The same
theorem 2.4 implies the convergence (4.10) of  the
bilinear form $\phi(t+T,{\bf x})$ to $\phi_{out}(t,{\bf
x})$ as $T\to +\infty.$

The constructed bilinear forms (4.5)-(4.7) are
sesquilinear.  Really, taking into account (4.13),  we
have that $$ \overline {u(t,{\bf x},z_1,z_2)} =
\sum\Bigl({\overline {u^{(l+1)}(t,{\bf x},z_1,z_2)} } -
{\overline {u^{(l)}(t,{\bf x},z_1,z_2)}}\Bigr) = u(t,{\bf
x},{\overline z}_2,{\overline z}_1),  $$ and so $$
\overline {u_{out}(t,{\bf x},z_1,z_2)} = u_{out}(t,{\bf
x},{\overline z}_2,{\overline z}_1).  $$ Hence for
$\chi_1 = \sum\alpha_j|v_j\rangle,$
$\chi_2=\sum\beta_k|w_k\rangle,$ $$ \overline{\phi(t,{\bf
x})(\chi_1,\chi_2)} = \phi(t,{\bf x})(\chi_2,\chi_1) ,$$
$$ \overline{:\!\phi^3(t,{\bf x})\!:(\chi_1,\chi_2)} =\;
 :\!\phi^3(t,{\bf x})\!:(\chi_2,\chi_1) ,$$ $$
\overline{\phi_{out}(t,{\bf x})(\chi_1,\chi_2)} =
\phi_{out}(t,{\bf x})(\chi_2,\chi_1) ,$$ i.e. bilinear
forms (4.5)-(4.7)  are sesquilinear.

Theorem 4.1 is proved.

\medskip The constructed bilinear form as a solution of
the  quantum  equation satisfies the uniqueness condition
of the  following type.  Let $$
\phi_1=\phi_{in}+N_R(\phi_1), $$ $$
\phi_2=\phi_{in}+N_R(\phi_2).  $$ Let $\phi_1$ and
 $\phi_2$ be bilinear forms defined on $D_\theta\times
D_\theta$  and such, that for $\chi_1,\chi_2\in D_\theta$
 \quad $\phi_1(t,{\bf x})(\chi_1,\chi_2)\in{\cal F}^C$
and $\phi_2(t,{\bf x})(\chi_1,\chi_2)\in{\cal F}^C.$ Let
$:\!\phi^3_1(t,{\bf x})\!:$ and $:\!\phi^3_2(t,{\bf
x})\!:$ be the bilinear forms and let the relation $$
:\!\phi^3(t,{\bf x})\!:  (|v\rangle,|w\rangle) =
 e^{-2({\bar v},w)}\Bigl(\phi(t,{\bf
x})(|v\rangle,|w\rangle) \Bigr)^3\eqno(4.19) $$ be
fulfilled for any pair of coherent vectors
$|v\rangle,|w\rangle$ (The definition of  normal
ordering!).  Then on $D_\theta \times D_\theta$ \quad
$\phi_1(t,{\bf x})$ coincides with $\phi_2(t,{\bf x})$
and with our form $\phi(t,{\bf x}).$

Here we do not consider the proof of the uniqueness and
this definition of a normal ordering. We use the relation
of the form (4.19) for a Wick  polynomial only.

\medskip
We formulate yet
two useful assertions.

\medskip
Theorem 4.2.

The bilinear forms $\phi(t,{\bf x}),$  $\phi_{out}(t,{\bf
 x})$ transform as scalar under the Poincar\'e
 transformation generated by the $in$-field $$
U_{in}(a,\Lambda)\phi(t,{\bf x})U_{in}(a,\Lambda)^{-1}
=\phi((a,\Lambda)(t,{\bf x})), \eqno(4.20)$$
$$U_{in}(a,\Lambda)\phi_{out}(t,{\bf
x})U_{in}(a,\Lambda)^{-1} =\phi_{out}((a,\Lambda) (t,{\bf
x})) \eqno(4.21)$$

\medskip
Remarks.

1. The  Poincar\'e transformation of the  interpolating
field and of the $in$- and $out$-fields generates the
same representation of the Poincar\'e group,
$U(a,\Lambda)=U_{in}(a,\Lambda)=U_{out}(a,\Lambda)$.

2. The expressions (4.20)-(4.21) are defined as bilinear
forms for all $a\in{I\!\!R}^4$ and $\Lambda$ such that
$U(0,\Lambda)D_\theta\in D_{2\theta}$, i.e. for $\Lambda$
sufficiently near to 1.  This is connected with the fact
that our bilinear form is defined only on $D_\theta\times
 D_\theta$ for sufficiently small $\theta$, and the norm
used for the space of initial $in$-data is
Lorentz-noninvariant.

3. It is possible to obtain analyticity of classical
solutions for small complex initial data for the space of
the initial data ${\cal F}^C$.  To derive this
analyticity one can use the estimates of Lemmas 2.1 and
2.2 and the method analogous to \cite{17} or given by
Baez and Zhou \cite{18} for the case of initial data,
corresponding to the finite energy.  This analyticity
allows to  extend  by  continuity the equalities (4.20)
 and (4.21)  on the Poincar\'e-invariant subspace.

\medskip Expressions (4.5), (4.7) of Theorem 4.1 for the
bilinear form $\phi$ and $\phi_{out}$ imply obviously
that the coupling constant $\lambda$ can be reconstructed
uniquely by matrix elements of the interpolating field
$\phi$ or the $out$-field $\phi_{out}$.  Moreover, the
following assertion is valid.

\medskip
Theorem 4.3.

The coupling constant $\lambda$ is determined uniquely
by  matrix elements (4.5) of the interpolating field
\begin{eqnarray*}
\lambda
&=&\lim_{\varepsilon\to 0}\varepsilon^{-4}\lim_{T\to
\infty}(\int(\phi(t,{\bf x})(|v\rangle,|v\rangle))^4
dt d^3 x)^{-1} \langle v|v\rangle^4 \\&&
\int(\phi(t+T,{\bf x})(|\varepsilon v
\rangle,|\varepsilon v\rangle)
\dot\phi(t+T,{\bf x})(|2\varepsilon v
\rangle,|2\varepsilon v\rangle)
\end{eqnarray*}
$$
-\phi(t+T,{\bf x})(|2\varepsilon v\rangle,
|2\varepsilon v\rangle)
\dot\phi(t+T,{\bf x})(|\varepsilon
v\rangle,|\varepsilon v\rangle)) d^3 x\eqno(4.22) $$
\begin{eqnarray*} &=&\lim_{\varepsilon\to
0}(\int(\phi(t,{\bf x})(|v\rangle,|v\rangle))^4
d t d^3 x)^{-1}\langle v|v\rangle^4
\\&& \int(\phi_{out}(t,{\bf x})
(|\varepsilon v\rangle,|\varepsilon v\rangle)
\dot\phi_{out}(t,{\bf x})(|2\varepsilon
v\rangle,|2\varepsilon v\rangle) \end{eqnarray*}
$$ -\dot\phi_{out}(t,{\bf x})(|2\varepsilon v
\rangle,|2\varepsilon v\rangle)
\phi_{out}(t,{\bf x}
(|\varepsilon v\rangle,|\varepsilon v\rangle))
d^3x\eqno(4.23)
$$
for any $|v\rangle\in D_\theta,$
 $v\in{\cal S}({I\!\!R}^3,C),$ $v\not=0.$

\medskip
Proof.

For
$v\in {\cal S}({I\!\!R}^3,C)$
the initial $in$-data
$u_{in}(t,{\bf x},\bar v, v)$
are real and belong to ${\cal F}$, and thus,
$u(t,{\bf x},\bar v,v)$ belongs to ${\cal F}$ also,
$u(t,{\bf x},\bar v,v)\not=0.$

Since
\begin{eqnarray*}
0&<&\int u(t,{\bf x},\bar v,v)^4dtd^3 x
\\&\leq&(\int\sup_{{\bf x}} u(t,{\bf x},\bar v,v)^2
dt)\sup_t\int u(t,{\bf x},\bar v,v)d^3 x
\\&\leq&\|u\|^4_F, \end{eqnarray*} so $(\int u(t,{\bf
x},\bar v,v)^4 dtd^3 x)^{-1}$ is correctly defined and,
by Theorem 4.1, is equal to $$ (\int(\phi(t,{\bf
x})(|v\rangle,|v\rangle))^4dtd^3 x)^{-1}\langle v | v
\rangle^4.  $$ The same theorem implies the existence of
the limit as $T\to +\infty$ and the right side
(4.22)=(4.23).  Taking into account that
$\langle\varepsilon v|\varepsilon v\rangle\to\langle
0|0\rangle=1 $ for $\varepsilon \to 0,$ the expressions
(4.22) and (4.23) are equal to $\lambda.$ This is the
consequence of the equality \begin{eqnarray*}
\lambda&=&\lim_{\varepsilon\to 0}\varepsilon^{-4} (\int
u(t,{\bf x},\bar v,v)^4 dtd^3 x)^{-1} \\&&
(\int(u_{out}(t,{\bf x},\varepsilon\bar v,\varepsilon
v)\dot u_{out} (t,{\bf x},2\varepsilon\bar v,
2\varepsilon v) \\&& -u_{out}(t,{\bf x},2\varepsilon\bar
v,2\varepsilon v)\dot u_{out} (t,{\bf x},\varepsilon\bar
v, \varepsilon v))d^3 x), \end{eqnarray*} which is proved
in \cite{15}. Theorem 4.3 is proved.

\medskip
Remarks.

1. It follows that the coupling constant is uniquely
defined by matrix elements of the $out$-field only.

2.  Expression (4.23) can be rewritten  also in the form
\begin{eqnarray*} \lambda&=&\lim_{\varepsilon\to
0}\varepsilon^{-4} (\int:\!\phi^4\!:(t,{\bf x})
(|v\rangle,|v\rangle)dtd^3 x)^{-1}\langle v|v\rangle \\&&
\int(\phi_{out}(t,{\bf x})(|\varepsilon
v\rangle,|\varepsilon v\rangle) \dot\phi_{out}(t,{\bf
x})(|2\varepsilon v\rangle,|2\varepsilon v\rangle) \\&&
-\phi_{out}(t,{\bf x})(|2\varepsilon
v\rangle,|2\varepsilon v\rangle) \dot\phi_{out}(t,{\bf
x})(|\varepsilon v\rangle,|\varepsilon v\rangle)) d^3 x,
\end{eqnarray*} where $:\!\phi^4(t,{\bf x})\!:$ is the
bilinear form with matrix elements (4.27).

\medskip
The proved assertion about bilinear forms implies easily
that on $D_\theta\times D_\theta$ the bilinear forms $$
\phi^{(l)}_{S,\tau,\Lambda}(t,{\bf x}), \quad
:\!(\dot\phi^{(l)2}_{S,\tau,\Lambda}(t,{\bf x})\!:, \quad
:\!(\phi^{(l)2}_{S,\tau,\Lambda}(t,{\bf x})\!:, \quad
:\!(\nabla\phi^{(l)}_{S,\tau,\Lambda})^2(t,{\bf x})\!:,
\quad :\!(\phi^{(l)4}_{S,\tau,\Lambda}(t,{\bf x})\!:, $$
$$ H^{(l)}_{S,\tau,\Lambda} ={1\over
2}\int\Bigl(:\!\dot\phi^{(l)2}_{S,\tau,\Lambda}(t,{\bf
x})\!:  + :\!(\nabla\phi^{(l)}_{S,\tau,\Lambda})^2(t,{\bf
 x})\!:  + m^2:\!\phi^{(l)2}_{S,\tau,\Lambda}(t,{\bf
 x})\!:  + {\lambda\over 2}
 :\!\phi^{(l)4}_{S,\tau,\Lambda}(t,{\bf x}\!:\Bigr)d^3x.
$$ are  correctly defined.  Clearly that as $l\to
\infty,$ $S\to -\infty,$ $\Lambda\to 1,$ and
$\sigma\to\delta$-function these bilinear form on
$D_\theta\times D_\theta$ converge to the bilinear forms
$$ :\!\dot\phi^2(t,{\bf x})\!:,\quad :\!\phi^2(t,{\bf
x})\!:,         \quad :\!(\nabla\phi)^2(t,{\bf
x})\!:,\quad :\!\phi^4(t,{\bf x})\!:, $$ $$ H={1\over
2}\int\Bigl( :\!\dot\phi^2(t,{\bf x})\!: +
:\!(\nabla\phi)^2(t,{\bf x})\!: + m^2:\!\phi^2(t,{\bf
 x})\!:  + {\lambda \over 2}:\!\phi^4(t,{\bf
 x})\!:\Bigr)d^3x.  $$ On $D_\theta\times D_\theta$
these bilinear forms satisfy the relations $$
:\!\dot\phi^2(t,{\bf x})\!:(\chi_1,\chi_2) =
\sum\overline{\alpha}_j\beta_k\langle v_j|w_k\rangle
\;\dot u(t,{\bf x},\overline{v}_j,w_k)^2, \eqno(4.24)$$
$$ :\!\phi^2(t,{\bf x})\!:(\chi_1,\chi_2) =
\sum\overline{\alpha}_j\beta_k\langle v_j|w_k\rangle
\;u(t,{\bf x},\overline {v}_j,w_k)^2, \eqno(4.25)$$ $$
:\!(\nabla\phi)^2(t,{\bf x})\!:(\chi_1,\chi_2) =
\sum\overline{\alpha}_j\beta_k\langle v_j|w_k\rangle
\;(\nabla u(t,{\bf x},\overline{ v}_j,w_k))^2,
\eqno(4.26)$$ $$ :\!\phi^4(t,{\bf x})\!:(\chi_1,\chi_2) =
\sum\overline{\alpha}_j\beta_k\langle v_j|w_k\rangle
\;u(t,{\bf x},\overline{ v}_j,w_k)^4, \eqno(4.27)$$ $$
H(\chi_1,\chi_2) = \sum\overline{\alpha}_j\beta_k\langle
 v_j|w_k\rangle $$ $$ {1 \over 2} \int\biggl(\dot
u(t,{\bf x},\overline{v}_j,w_k)^2 + (\nabla u(t,{\bf
 x},\overline{ v}_j,w_k))^2 + m^2 u(t,{\bf x},\overline
{v}_j,w_k)^2 $$ $$ + {\lambda\over 2} u(t,{\bf
x},\overline{ v}_j,w_k)^4\biggr)d^3x.  \eqno(4.28)$$
Moreover, on  $D_\theta \times D_\theta$ the bilinear
form $H$ is equal to the bilinear form $H_{in},$ $$
H(\chi_1,\chi_2) = H_{in}(\chi_1,\chi_2) =
\sum\overline{\alpha}_j\beta_k \langle v_j|w_k\rangle $$
$$ {1\over 2}\int\biggl(\dot u_{in}(t,{\bf x}, \overline{
v}_j,w_k)^2 + (\nabla u_{in}(t,{\bf x},\overline{
v}_j,w_k))^2 + m^2 u_{in}(t,{\bf
 x},\overline{v}_j,w_k)^2\biggr)d^3x $$
$$=\sum\overline{\alpha}_j\beta_k\langle v_j|w_k\rangle
\int\mu(k) \overline{v\tilde{ }_j(k)} w_k(k) d^3k \biggr)
.  $$ Therefore, on $D_\theta \times D_\theta$ the
bilinear form $H(\chi_1,\chi_2)$ is positively definite
and $H_{in}$ is a unique positive self-adjoint operator,
which bilinear form on $D_\theta \times D_\theta$
 coincides with the bilinear form of $H.$

We remark, that the expressions analogous to (4.24)-(4.28)
can be written for the momentum and
 angular momentum  operators.

\section{Discussion}

We mention first the review of
Callaway \cite{24}. This review contains, in
particular,    arguments  of Fr\"ohlich \cite{25}
and Aizenman et al. \cite{26,27}
concerning the triviality of $\phi^4_4.$
Fr\"ohlich \cite{25} and Aizenman et al. \cite{26,27}
assert that any construction of $\phi^4_4,$ obtained as a limit
of ferromagnetic lattice approximation, is trivial.
Recently Pedersen,
Segal, Zhou \cite{28} gave
arguments for nontriviality of (massless) $\phi^4_4,$
and, more generally, for nontriviality
of $\phi^q_d$) \cite{28},
see also \cite{29,30}. The usual perturbation theory
claims that $\phi^4_4$ is a nontrivial and a
renormalizable theory.

We interpret the results of
Fr\"ohlich \cite{25} and Aizenman et al. \cite{26,27}
as an approximation of a measure.
This measure corresponds to the approximation, but this
 approximation do not catch the nonlinearity (and
 singularity) of the interaction. Its convergence
corresponds to the convergence on the subspace of zero
measure (for the true measure), see the analogous
interpretation for more singular case \cite{31,32,33}.

Our approach obtains undoubtedly a nontrivial theory, in
particular, the coupling constant is uniquely determined
by the matrix elements of the interpolating or the
out-field (see Theorem 4.3), but the straightforward
proof is up to now unknown and absent. It would be very
interesting to understand to what structure corresponds
our construction: to the whole quantum field or only to
the non-linear ``tree" approximation.

This construction is connected with the idea to construct
the vacuum with the help of a  generalized density and/or
 its logarithmical derivative, see \cite{34}-\cite{41}.
In our case the generalized density is equal to
$\rho_{in}(w u),$ i.e. to the vacuum in terms of the
interacting field, it is defined on the (whole) space
${\cal F}.$ Here $\rho_{in}(\cdot)$ is the generalized
density of the free vacuum and $w$ is a (quasi) canonical
transformation.  $\rho_{in}(w u)$ has to consider on
finite-dimensional subspaces and then has to be extended
on ${\cal S}'$ (as a measure generated by this
generalized density).  To $\rho_{in}(\cdot)$ corresponds
a unique state. This state can be obtained by extension
from finite-dimensional subspaces of entire holomorphic
functions \cite{11,34,35}.  In our case the variable
 $\varphi(x)-i((-\Delta_3+m^2)^{-1/2}\pi)(x)$ corresponds
to the complex variable. Some exceptional properties of
the state in terms of complex variables are described in
\cite{11}.  They correspond to a holomorphic
representation of Weyl group. In this case the Weyl group
is a nuclear infinite-dimensional Lie group.

The further progress will be connected with the
possibility to extend a domain of definition of the
bilinear form and/or with possibility to obtain the
bounds connected with these bilinear forms in some
 suitable rigging of Fock Hilbert space of the
$in$-field.  It is very important that the Hamiltonian is
correctly defined as an operator, essentially
self-adjoint on $D_\theta$, and in the same time it can
be expressed as a bilinear form connected with $\phi.$ We
emphasize that we need such bounds in terms of bilinear
form corresponding to the field $\phi.$

\medskip
\medskip
{\large Acknowledgments}

The author is grateful to Ludwig Faddeev for the
interesting question.

\end{document}